\documentclass[12pt]{article}
\usepackage{amsfonts,epsfig,psfig,cite}

\begin{document}

%% \date{February 26, 2002}

\def\reff#1{(\ref{#1})}
\newcommand{\be}{\begin{equation}}
\newcommand{\ee}{\end{equation}}
\newcommand{\<}{\langle}
\renewcommand{\>}{\rangle}

%%%  \ltapprox and \gtapprox produce > and < signs with twiddle underneath
\def\spose#1{\hbox to 0pt{#1\hss}}
\def\ltapprox{\mathrel{\spose{\lower 3pt\hbox{$\mathchar"218$}}
 \raise 2.0pt\hbox{$\mathchar"13C$}}}
\def\gtapprox{\mathrel{\spose{\lower 3pt\hbox{$\mathchar"218$}}
 \raise 2.0pt\hbox{$\mathchar"13E$}}}

\def\bsigma{\mbox{\protect\boldmath $\sigma$}}
\def\bpi{\mbox{\protect\boldmath $\pi$}}
\def\smfrac#1#2{{\textstyle\frac{#1}{#2}}}
\def\smhalf{ {\smfrac{1}{2}} }

\newcommand{\re}{\mathop{\rm Re}\nolimits}
\newcommand{\im}{\mathop{\rm Im}\nolimits}
\newcommand{\tr}{\mathop{\rm tr}\nolimits}
\newcommand{\fr}{\frac}
\newcommand{\diti}{\frac{\mathrm{d}^2t}{(2 \pi)^2}}

\def\Z{{\mathbb Z}}
\def\R{{\mathbb R}}
\def\C{{\mathbb C}}

\title{Two-Dimensional Heisenberg Model with Nonlinear Interactions:
       $1/N$ Corrections}
\author{
  \\[-0.5cm]
  {\small Sergio Caracciolo}              \\[-0.2cm]
  {\small\it Dipartimento di Fisica dell'Universit\`a di Milano, 
           I-20133 Milano} \\[-0.2cm]
  {\small\it INFN, Sez. di Milano, and INFM-NEST, Pisa, ITALY}    \\[-0.2cm]
  {\small {\tt Sergio.Caracciolo@sns.it}}     \\[-0.2cm]
  \\[-0.35cm] \and
  {\small Bortolo Matteo Mognetti}              \\[-0.2cm]
  {\small\it Dipartimento di Fisica and INFN -- Sezione di Milano} \\[-0.2cm]
  {\small\it  Universit\`a di Milano, I-20133 Milano, ITALY} \\[-0.2cm]
  {\small {\tt Bortolo.Mognetti@mi.infn.it}}     \\[-0.2cm]
  \\[-0.35cm] \and
  {\small Andrea Pelissetto }        \\[-0.2cm]
  {\small\it Dipartimento di Fisica and INFN -- Sezione di Roma I}  \\[-0.2cm]
  {\small\it Universit\`a degli Studi di Roma ``La Sapienza''}      \\[-0.2cm]
  {\small\it I-00185 Roma, ITALY}        \\[-0.2cm]
  {\small {\tt Andrea.Pelissetto@roma1.infn.it}}  \\[-0.2cm]
  {\protect\makebox[5in]{\quad}}  % To force authors' names to be written
                                  %   vertically, one above another.
                                  % (\author seems to put them side-by-side
                                  %   if there is room.)
   \\
}

\vspace{0.2cm}
\maketitle
\thispagestyle{empty}   % Suppress page number on front page.

\vspace{-0.4cm}

\begin{abstract}
We investigate a two-dimensional classical $N$-vector model 
with a generic nearest-neighbor interaction $W(\bsigma_i\cdot \bsigma_j)$
in the large-$N$ limit, focusing on the finite-temperature transition point at
which energy-energy correlations become critical. We show 
that this transition belongs to the Ising universality class. 
However, the width of the region in which Ising behavior is observed 
scales as $1/N^{3/2}$ along the magnetic direction and as $1/N$ in the 
thermal direction; outside a crossover to mean-field behavior occurs.
This explains why only mean-field behavior is observed for $N=\infty$.
\bigskip 


\end{abstract}

\clearpage

%%%% \doublespace

\section{Introduction} 

The two-dimensional Heisenberg model has been the object of 
extensive studies which mainly focused on the $O(N)$-symmetric
Hamiltonian
\be
H = - N \beta \sum_{\<ij\>} \bsigma_i\cdot \bsigma_j,
\label{NLSigma}
\ee
where $\bsigma_i$ is an $N$-dimensional unit spin and 
the sum is extended over all lattice nearest neighbors.
The behavior of this system in two dimensions is well
understood. It is disordered for all finite $\beta$
\cite{MW-66} and it is described for $\beta\to\infty$ 
by the perturbative renormalization group 
\cite{Polyakov-75,BZ-76,BLS-76}. The square-lattice
model has been extensively studied numerically
\cite{Wolff_89_90,EFGS-92,Kim,CEPS-95,CEMPS-96,MPS-96},
checking the perturbative predictions 
\cite{FT-86,CP-94,CP-95,ABC-97}
and the nonperturbative constants 
\cite{HMN-90,HN-90,CPRV-97}.

Beside Hamiltonian (\ref{NLSigma}), one can also consider more general
interactions. In particular, one can consider the most general Hamiltonian
with nearest-neighbor couplings given by
\be
H = - N\beta \sum_{\<ij\>} W(1 + \bsigma_i\cdot \bsigma_j),
\label{Hgenerica}
\ee
where $W(x)$ is a generic function such that
$W(2) > W(x)$ for all $0\le x < 2$, in order to
guarantee that the system orders ferromagnetically
for $\beta\to\infty$. The
behavior of model \reff{Hgenerica} for $\beta\to \infty$ is similar to that of 
model (\ref{NLSigma}): the critical behavior is described 
by the perturbative renormalization group, the spin-spin correlation length
diverging as $\beta\to \infty$. 

While for $\beta\to \infty$ the behavior is similar, 
for finite values of $\beta$ the more general Hamiltonian 
\reff{Hgenerica} may give rise to a more complex phase diagram.
In particular, one may observe first-order transitions, as it has been 
shown rigorously for a large class of functions 
$W(x)$ in Refs.~\cite{VES-02,VES-03}.  
These rigorous results confirm the conclusions of 
Refs.~\cite{MR-87,BGH-02,CP-02,PR-03} that found first-order transitions 
in the mixed O($N$)-$RP^{N-1}$ for $N=\infty$, 
in a modified $RP^{N-1}$ 
for $N=3$, and in systems 
with $W(x) \sim x^p$ for $p$ large enough for $N=3$ and $N=\infty$ 
respectively. The presence of first-order transitions does not violate
the Mermin-Wagner theorem \cite{MW-66}, since the spin-spin 
correlation length is finite at the transition.  However, the model studied in 
Ref.~\cite{MR-87} for $N=\infty$ also shows a finite-$\beta$ critical point.
The specific heat diverges indicating that energy fluctuations become critical.
On the other hand, the spin-spin correlation length is finite, although
nonanalytic, at the transition, so that the Mermin-Wagner theorem \cite{MW-66}
is not violated.
For a long time it has not been clear whether the observed critical point was 
an artifact of the large-$N$ limit or rather a true transition 
to be expected also for finite values of $N$
\cite{MR-87,Niedermayer_96,SS-01}. 
Recently, Bl\"ote, Guo, and Hilhorst \cite{BGH-02} 
studied numerically a class of systems 
with $W(x) \sim x^p$ and $N=3$ and found for a specific value of 
$p$ a critical transition with essentially the same features as those
observed in Ref.~\cite{MR-87} for $N=\infty$. This confirms the 
existence of finite-$\beta$ critical points where energy-energy 
correlations are critical, while the spin-spin correlation function is 
still short-ranged.

What remains to be understood is the nature of the critical point. 
A simple symmetry argument \cite{BGH-02} 
indicates that the transition should belong to the 
Ising universality class for any value of $N$, 
and this is confirmed by the numerical results 
for $N=3$. On the other hand, the large-$N$ analysis \cite{MR-87,CP-02}
finds mean-field exponents. In this paper we wish to reconcile these 
two apparently contradictory results. Indeed, we show that, as soon as 
large-$N$ fluctuations are considered, the transition becomes an Ising one. 
The mechanism that works in the large-$N$ limit is very similar to that 
observed in medium-range models 
\cite{crossover1,crossover2,crossover3,PRV-99,crossover4}, 
with $N$ playing the role of the 
interaction range. If $t \equiv (\beta - \beta_c(N))/\beta_c(N)$ is the reduced 
temperature and $h$ is a properly defined scaling field that plays the role 
of the Ising magnetic field, we find that a universal scaling behavior 
is obtained by taking $t\to 0$, $h\to 0$, $N\to \infty$ and 
keeping fixed $\tilde{t} \equiv t N$, $\tilde{h} \equiv h N^{3/2}$. 
Ising behavior is observed only for $\tilde{t}\ll 1$ and $\tilde{h} \ll 1$, 
so that the width of the Ising critical region goes to zero as 
$N^{-3/2}$ (except in the thermal direction where it goes to zero as 
$N^{-1}$) and no Ising behavior is observed for $N=\infty$. 
In the other limit mean-field behavior is observed instead. 

The analysis of the large-$N$ corrections is quite complex.
As in the standard $1/N$ expansion we first derive the effective Hamiltonian
for the auxiliary fields. However, at the critical point the 
propagator is singular, forbidding a standard $1/N$ expansion. 
We single out the zero mode and show that the effective Hamiltonian
that is obtained by integrating the massive modes corresponds to 
a weakly coupled $\phi^4$ theory. The analysis is somewhat complicated 
by the fact that the $\mathbb{Z}_2$ symmetry is absent and thus there is 
no natural definition of the thermal and magnetic scaling fields. The correct 
scaling fields are linear combinations of the Hamiltonian parameters that 
must be computed order by order in $1/N$. A detailed field-theoretical analysis 
of the effective Hamiltonian allows us to compute the $1/N$ corrections 
to the position of the critical point and to express the large-$N$ 
scaling behavior close to the transition in terms of Ising crossover 
functions.

The paper is organized as follows. In Sec.~\ref{sec2} we review the large-$N$
limit for model (\ref{Hgenerica}) on a square lattice, 
reporting some basic formulas 
needed in the paper. Derivations are reported in Ref.~\cite{CP-02}. 
In Sec.~\ref{sec3} we determine the linear scaling fields for $N=\infty$ 
and show that they are related by the mean-field equation of state. 
In Sec.~\ref{sec4} and \ref{sec5} we determine the effective Hamiltonian
for the auxiliary fields and for the zero mode. Sec.~\ref{sec6} 
discusses the weakly coupled $\phi^4$ theory in the absence of 
$\mathbb{Z}_2$ symmetry, determining the relevant renormalization constants
and the scaling behavior. Finally, in Sec.~\ref{sec7} we apply these results 
to the large-$N$ limit. We determine the correct scaling fields, the $1/N$ 
corrections to the position of the critical point and the width of the 
Ising critical region. In Sec.~\ref{sec8} we explicitly give the scaling 
behavior for $\< \bsigma_x\bsigma_{x+\mu}\>$, so that we can compare with 
numerical results, for instance with those of Ref.~\cite{BGH-02}. 
Finally, in Sec.~\ref{sec9} we report some numerical results for the 
mixed O($N$)-$RP^{N-1}$ discussed in Ref.~\cite{MR-87} and 
for the Hamiltonian introduced in Ref.~\cite{BGH-02}. 
Conclusions are presented in Sec.~\ref{sec10}.

\section{The large-$N$ limit} \label{sec2}

We consider Hamiltonian \reff{Hgenerica} on a square lattice and normalize 
$W(x)$ by requiring $W'(2)=1$ so that in the spin-wave limit 
\be
H = {N\beta\over 2} \int dx\, \partial_\mu\bsigma\cdot \partial_\mu\bsigma.
\ee 
The standard $N$-vector model corresponds to $W(x) = x$. 
In this work we do not specify the function $W(x)$; we only assume 
that it depends on a parameter $p$ and that in the $(\beta,p)$
plane there is a first-order transition line 
$\beta = \beta_f(p)$ ending at a critical point $(\beta_c,p_c)$ that 
will be characterized below.

Our discussion strictly applies only to ferromagnetic models, but under 
standard assumptions can be extended to $RP^{N-1}$ models
which correspond to functions satisfying $W(1+x) = W(1-x)$. In this 
case the Hamiltonian is invariant under the local transformations
$\bsigma_x \to z_x \bsigma_x$, $z_x = \pm 1$. 

The large-$N$ limit has been discussed in detail in Ref.~\cite{CP-02}.
We introduce three auxiliary fields 
$\lambda_{x\mu}$, $\rho_{x\mu}$, and $\mu_x$ in order to linearize the 
dependence of the Hamiltonian on the spins $\bsigma$ and 
to eliminate the constraint $\bsigma_x^2 = 1$. 
The partition function  becomes
\be
Z = \int \prod_{x\mu} [d\rho_{x\mu} d \lambda_{x\mu}]
         \prod_x [d\mu_x d\bsigma_x]\,  e^{-NA},
\ee
where 
\be
A = -{\beta\over2} \sum_{x\mu} 
   \left[\lambda_{x\mu} + \lambda_{x\mu}\bsigma_x\cdot\bsigma_{x+\mu} - 
         \lambda_{x\mu} \rho_{x\mu} + 2 W(\rho_{x\mu})\right] +
   {\beta\over2} \sum_{x} 
     \left(\mu_x\bsigma_x^2 - \mu_x\right).
\ee
We perform a saddle-point integration by writing
\begin{eqnarray}
\lambda_{x\mu} &=& \alpha + {1\over \sqrt{N}} 
   \widehat{\lambda}_{x\mu}, \nonumber \\
\rho_{x\mu} &=& \tau + {1\over \sqrt{N}} \widehat{\rho}_{x\mu}, \nonumber \\
\mu_{x} &=& \gamma + {1\over \sqrt{N}} \widehat{\mu}_{x}. 
\label{field-exp}
\end{eqnarray}
In Ref.~\cite{CP-02} we computed the corresponding saddle-point equations.
They can be written as
\begin{eqnarray}
\gamma &=& \alpha (4 + m_0^2)/2, \\
\alpha &=& 2 W'(\tau), \\
{\tau} &= &
   \overline{\tau}\left(m_0^{2}\right) \equiv 2 + {m_0^2\over 4} - {1\over 4 B_1(m_0^2)},
\label{def-taubar} \\
\beta &=& {B_1(m_0^2)\over W'(\overline{\tau},p)},
\label{puntosella2}
\end{eqnarray}
where the parameter $m_0$ is related to the spin-spin correlation  length
$\xi_\sigma = 1/m_0$ and
\begin{equation}
  B_n(m_0^2) \equiv \int_\mathbf{q}
   {1\over (\hat{q}^2 + m_0^2)^n},
\label{defBn}
\end{equation}
where the integral is extended over the first Brillouin zone.

In Ref.~\cite{CP-02} it was shown that generic models may show first-order
transitions. This happens when, for given $\beta$, there are several
values of $m_0^2$ that solve the gap equation (\ref{puntosella2}). 
Here, we will be interested at the endpoint of the first-order transition
line $m_0 = m_{0c}$, $p = p_c$.  At this point we have 
\begin{equation}
{\partial \beta\over \partial m_0^2} = 0, \qquad 
{\partial^2 \beta\over \partial (m_0^2)^2} = 0,
\label{eq-punto-critico}
\end{equation}
where the derivatives are taken at fixed $p$. 
We will only consider the generic case, always assuming that 
the third derivative is nonvanishing at the transition.

\section{Equation of state and scaling fields} \label{sec3}

We wish now to parametrize the singular behavior for $\beta \to \beta_c$ 
and $p \to p_c$. Expanding the gap equation (\ref{puntosella2})
near the critical point
we obtain
\begin{equation}
\beta - \beta_c = \sum_{nm} a_{nm} (p-p_c)^n (m_0^2 - m_{0c}^2)^m.
\label{expansion-gap}
\end{equation}
Because of the definition of $\beta_c$ we have $a_{00} = 0$. 
Moreover, Eq.~(\ref{eq-punto-critico}) implies
that $a_{01} = a_{02} = 0$. For $p=p_c$ we see that $m_0^2$ has the leading 
behavior 
\begin{equation}
    m_0^2 - m_{0c}^2 \approx 
   \left( {\beta - \beta_c \over a_{03}}\right)^{1/3},
\end{equation}
while for $\beta = \beta_c$, we have 
\begin{equation}
    m_0^2 - m_{0c}^2 \approx 
   \left( -{a_{10} (p-p_c) \over a_{03}}\right)^{1/3}.
\end{equation} 
The nonanalytic behavior with exponent 1/3 is observed along any straight 
line approaching the critical point, except that satisfying 
$\beta - \beta_c - a_{10} (p-p_c) = 0$. Therefore, the correct 
linear scaling field is 
\begin{equation}
  u_h \equiv \beta - \beta_c - a_{10} (p-p_c),
\label{scalingfield-uh}
\end{equation}
and we have $m_0^2 - m_{0c}^2 \sim u_h^{1/3}$ whenever $u_h\not = 0$. 
To include the case $u_h = 0$, we write the general scaling equation
\begin{equation}
m_0^2 - m_{0c}^2 = u_h^{1/3} f(x) ,
\label{m0-scaling}
\end{equation}
where $f(x)$ is a scaling function and $x$ is a scaling variable to be 
determined. Then, we use again the gap equation (\ref{expansion-gap}). 
Keeping only the leading terms we obtain 
\begin{equation}
u_h = a_{20} (p-p_c)^2 + a_{11}(p-p_c) u_h^{1/3} f(x) + 
      a_{03} u_h f(x)^3,
\end{equation}
so that 
\begin{equation} 
   a_{03} f(x)^3 + a_{11}(p-p_c) u_h^{-2/3} f(x) - a_{20} (p-p_c)^2 u_h^{-1} 
    - 1 = 0
\end{equation}
Since $(p-p_c)^2 u_h^{-1} = [(p-p_c) u_h^{-2/3}]^2 u_h^{1/3}$, 
the third term can be neglected. Thus, we may take
(the prefactor has been introduced for later convenience)
\begin{equation}
x \equiv a_{11} (p-p_c) |u_h|^{-2/3}. 
\end{equation}
The function $f(x)$ satisfies 
\begin{equation}
a_{03} f(x)^3 + x f(x) - 1 = 0.
\label{eq-for-f}
\end{equation}
Such an equation is exactly the mean-field equation for the magnetization.
Indeed, if we consider the mean-field Hamiltonian 
\begin{equation} 
{\cal H} = - h M + {t\over2} M^2 + {u\over 24} M^4,
\end{equation}
the stationarity condition gives
\begin{equation}
- h + t M + {u\over6} M^3 = 0,
\end{equation}
which is solved by $M = h^{1/3} \hat{f}(t |h|^{-2/3})$, where 
$\hat{f}(x)$ satisfies Eq.~(\ref{eq-for-f}) with $a_{03} = u/6$. 
It is thus clear that the scaling field (\ref{scalingfield-uh}) corresponds 
to $h$, while $p$ corresponds to the temperature. Note that this 
identification is not unique, since only the line $H=0$ is 
uniquely defined by the singular behavior.
For instance, in the usual Ising case,
we could define $t' = t + a h$ without changing the scaling equation of state,
since $t |h|^{-2/3} = t' |h|^{-2/3} + a h^{1/3}$. Since the scaling 
limit is taken with $h\to0$, $t\to 0$ at fixed $t |h|^{-2/3}$, we see that 
$t |h|^{-2/3} \approx t' |h|^{-2/3}$. In the Ising case, however, 
there is exact $\mathbb{Z}_2$ symmetry and thus the natural
$t$ variable is defined so that it is invariant under the symmetry.
In our case we could define $u_t = p - p_c + A (\beta - \beta_c)$
and fix $A$ by requiring the leading correction on any line (except $u_h = 0$)
to be of order $u_h^{2/3}$ instead of order $u_h^{1/3}$, recovering in this way
an approximate $\mathbb{Z}_2$ symmetry. For our purposes this is 
irrelevant and thus we will use $p-p_c$ as thermal scaling field.

Eq.~(\ref{eq-for-f}) may have more than one solution. A simple 
analysis shows that, if $a_{03} > 0$, 
$f(x)$ is the unique positive solution of Eq.~(\ref{eq-for-f}) and 
has the following asymptotic behaviors: 
$f(x) \approx \sqrt{-x/a_{03}}$ for $x\to -\infty$ and
$f(x) \approx 1/x$ for $x\to \infty$. If $a_{03}$ is negative 
the same results apply to $-f(-x)$. 

Eq.~(\ref{m0-scaling}) gives the leading behavior. It is also possible to 
compute the subleading corrections. We obtain for the leading one
\begin{equation}
m_0^2 - m_{0c}^2 = u_h^{1/3} f(x) + u_h^{2/3} g(x) + O(u_h),
\end{equation}
with 
\begin{equation}
g(x) = - {a_{20} x^2 + a_{12} a_{11} x f(x)^2 + a_{04} a_{11}^2 f(x)^4 \over 
        a_{11}^2 (x + 3 a_{03} f(x)^2)}.
\end{equation}
Finally, let us discuss the singular behavior of the energy.
We have 
\begin{eqnarray} 
    E = 2 W(\overline{\tau},p).
\end{eqnarray}
Such a function is regular in $m_0^2$ and $p$. Since 
$p - p_c\sim x |u_h|^{2/3}$ and $m_0^2 - m_{0c}^2 \sim u_h^{1/3}$, 
the leading term is obtained by expanding the previous equation in powers of 
$m_0^2 - m_{0c}^2$. Thus
\begin{equation}
   E = 2 W_c + 2 W'_c \left.{B_1^2 - B_2\over 4 B_1^2}\right|_{m_0 = m_{0c}}
    u_h^{1/3} f(x)  + O(u_h^{2/3}).
\end{equation}
where $W' = \partial W(x,p)/\partial x$ with $x = \overline{\tau}$ 
and the suffix $c$ indicates that all quantities must be computed at the 
critical point.

\section{The $1/N$ calculation: propagator and effective vertices} \label{sec4}

In order to perform the $1/N$ calculation we integrate out the 
fields $\bsigma$, which is straightforward since the 
Hamiltonian is quadratic in these fields. We thus obtain 
\be
Z = \int \prod_{x\mu} [d\rho_{x\mu} d \lambda_{x\mu}]
         \prod_x d\mu_x\,  e^{-{\cal H}}. 
\ee
where the effective Hamiltonian ${\cal H}$ can be expanded in powers of the 
auxiliary fields. If we define a five-component field $\Psi_A$
\begin{equation}
\Psi = (\hat{\mu}, \hat{\lambda}_1, \hat{\lambda}_2,
        \hat{\rho}_1, \hat{\rho}_2),
\end{equation}
then ${\cal H}$ can be written as 
\begin{eqnarray}
{\cal H} &=& {1\over 2} \int_{\mathbf{p}} \sum_{A_1A_2} 
   \Psi_{A_1}(-\mathbf{p}) P_{A_1A_2}^{-1}(\mathbf{p}) \Psi_{A_2}(\mathbf{p})
\label{effexp} \\
  &+& \sum_{n=3} {1\over n!} {1\over N^{n/2-1} }
  \int_{\mathbf{p}_1} \cdots \int_{\mathbf{p}_n} 
    \delta\left(\sum_i \mathbf{p}_i\right) \; 
   \sum_{A_1,\ldots,A_n} 
   V^{(n)}_{A_1,\ldots,A_n} (\mathbf{p}_1,\ldots,\mathbf{p}_n)
   \Psi_{A_1}(\mathbf{p}_1)\cdots \Psi_{A_n}(\mathbf{p}_n),
\nonumber 
\end{eqnarray}
where the indices $A_i$ run from 1 to 5, 
\be
  \int_\mathbf{p} \equiv \int {\mathrm{d}^2 p\over (2\pi)^2},\qquad \delta\left(\mathbf{p}\right)  \equiv \prod_{\alpha=1}^{2} 2 \pi \delta\left(p_{i}\right),
\ee
and the integration is over the first Brillouin zone. 
The propagator can be explicitly written as\footnote{
It is useful to write the Fourier transform of $\hat{\lambda}_x$ as 
$\hat{\lambda}(\mathbf{p}) = 
e^{-i p_\mu/2} \sum_x e^{-ip\cdot x} \hat{\lambda}_x$. This makes all vertices 
and propagators real.}
\begin{equation}
\mathbf{P^{-1}(p)}= - {1\over {W'}^2}
\left( \begin{array}{ccccc}
\fr{1}{2}A_{0,0} &-\fr{1}{2} A_{1,0} &-\fr{1}{2}A_{0,1} & 0 & 0 \\
-\fr{1}{2}A_{1,0} &\fr{1}{2} A_{2,0} &\fr{1}{2}A_{1,1} &
-\frac {1}{2}\beta {W'}^2 & 0 \\
-\fr{1}{2}A_{0,1} & \fr{1}{2}A_{1,1} &\fr{1}{2} A_{0,2} & 0 &
 -\frac{1}{2}\beta {W'}^2\\
0 & - \frac{1}{2}\beta {W'}^2 & 0 & \beta W'' {W'}^2 & 0 \\
0 & 0 & - \frac{1}{2}\beta {W'}^2 & 0 & \beta W''{W'}^2
\end{array} \right),
\label{propagator}
\end{equation} 
where $W$ should always be intended as a function of $\bar{\tau}(m_0^2)$,
\begin{equation}
A_{i,j}(\mathbf{p},m_0^2) \equiv \int_\mathbf{q}
\frac{  \cos^i q_x \cos^j q_y }{
     [m_{0}^2 + \widehat{(q + \frac{p}{2})}^2 ]
     [m_{0}^2 + \widehat{(q - \frac{p}{2})}^2]  },
\end{equation}
and $\hat{p}^2 \equiv 4 (\sin^2 p_x/2 + \sin^2 p_y/2)$.  

For $p\to0$, by using the algebraic algorithm described in App. A of
Ref.~\cite{CP-95}, it is easy to express the integrals $A_{i,j}(0,m_{0}^{2})$
in terms of the integrals $B_n(m_0^2)$ defined in Eq.~\reff{defBn} 
with $n=1,2$. Explicitly we have
\begin{eqnarray}
&& A_{00}(\mathbf{0},m_0^2) = B_2,
\nonumber \\
&& A_{10}(\mathbf{0},m_0^2) = A_{01}(\mathbf{0},m_0^2) =
    \left(1 + {m_0^2\over 4}\right) B_2 - {1\over4} B_1 ,
\nonumber \\  
&& A_{11}(\mathbf{0},m_0^2) =
  - {1\over8}(4+m_0^2) B_1 + {1\over8} (8 + 8 m_0^2 + m_0^4) B_2,
\nonumber \\
&& A_{20}(\mathbf{0},m_0^2) = A_{02}(\mathbf{0},m_0^2) =
 {1\over8} + B_2 - {1\over8}(4 + m_0^2) B_1.
\label{relAijBn}
\end{eqnarray}
Vertices are analogously computed. It is easy to check that the only 
nonvanishing contributions for which some $A_i$ is equal to 4 or 5 are 
\be
V_{4\ldots 4}^{(n)}(\mathbf{p}_1,\ldots,\mathbf{p}_n) =  
V_{5\ldots 5}^{(n)}(\mathbf{p}_1,\ldots,\mathbf{p}_n) =  
   - \beta W^{(n)}(\bar{\tau}).
\ee
If all indices satisfy $A_i\le 3$, then
\begin{eqnarray}
&& V_{A_1,\ldots,A_n}^{(n)}(\mathbf{p}_1,\ldots,\mathbf{p}_n) 
\delta(\sum_i \mathbf{p}_i) = 
\label{defVn}
\\
&& = 
 {(-1)^{n+1} \over [W'(\bar{\tau})]^n} \left\{
  \prod_{i=1}^n \left[ \int_{\mathbf{q}_i} 
\delta(\mathbf{q}_{i+1} - \mathbf{q}_i - \mathbf{p}_i) 
   {1\over \hat{q}_i^2 + m_0^2} R_{A_i} (\mathbf{p}_i,\mathbf{q}_i)
    \right] + {\rm permutations}\right\},
\nonumber
\end{eqnarray}
where
\be
R_1(\mathbf{p},\mathbf{q}) = 1, \qquad 
R_2(\mathbf{p},\mathbf{q}) = - \cos(q_{x} + p_{x}/2), \qquad 
R_3(\mathbf{p},\mathbf{q}) = - \cos(q_{y} + p_{y}/2).
\ee
The permutations should made the quantity in braces 
symmetric under any exchange of $(\mathbf{p}_i,A_i)$ (the total number 
of needed terms is $(n-1)!/2\,$). 
As already discussed in Ref.~\cite{CP-02}, at the critical point 
the inverse propagator at zero momentum has a vanishing eigenvalue. 
Indeed, a straightforward computation gives 
\begin{eqnarray}
{\rm det}\, \mathbf{P}^{-1} (\mathbf{0}) &=& K_{0,\rm det} s_1,
\label{detPinv-p0}
\end{eqnarray}
where $K_{0,\rm det}$ is given by 
\begin{equation}
 K_{0,\rm det}\equiv
  - {B_1^3\over 128 {W'}^6}
  \left[4 B_1 W' - (1 - m_0^2 (8 + m_0^2) B_2) W''\right],
\end{equation}
and 
\begin{equation}
s_1 \equiv - {\partial \beta \over \partial m_0^2} =
{1\over 4 B_1 {W'}^2} (4 B_1 B_2 W' + B_1^2 W'' - B_2 W'').
\label{dbdm}
\end{equation}
Eq.~(\ref{detPinv-p0}) shows that the determinant vanishes at the
critical point---hence there is at least one vanishing eigenvalue---since 
there $\partial \beta/\partial m_0^2 = 0$. 
The corresponding eigenvector  can be written as 
\begin{equation}
\mathbf{z} = \left(2 {A_{01}(\mathbf{0},m_0^2)\over A_{00}(\mathbf{0},m_0^2)}, 
          1, 1, {1\over 2 W''},
          {1\over 2 W''} \right)
\label{defz}
\end{equation} 
computed at the critical point. Indeed,
\begin{equation}
\sum_{B = 1}^5 (P^{-1})_{AB}(\mathbf{0}) z_B  =
        {B_1 \over 4 B_2 W''} s_1 (0,1,1,0,0)_{A}.
\label{eqZ}
\end{equation}
Note that there is always only one zero mode. Indeed, at the critical 
point, we have from Eq.~\reff{dbdm} 
\be
W'' = - {4 B_1 B_2 \over B_1^2 - B_2} W',
\ee
so that we can write at criticality
\be
 K_{0,\rm det} = - {B_1^4 [B_1^2 - (8 + m_0^2) m_0^2 B_2^2]\over 
                    32 (B_1^2 - B_2)} {1\over {W'}^5}.
\ee
We have verified numerically that the prefactor of $1/{W'}^5$ is always finite
and negative, so that $K_{0,\rm det}$ is always nonvanishing. 
Thus, for $p = p_c$ the determinant
${\rm det}\, \mathbf{P}^{-1} (\mathbf{0}) $ vanishes 
as $(m_0^2 - m_0^2)^2$, as the eigenvalue associated with the zero mode,
cf.~Eq.~\reff{eqZ}. Thus, there can only be a single eigenvector with zero
eigenvalue.

Because of the zero mode, it is natural to express the fields in terms of a new
basis. For each $m_0^2$ and $p$,
given the inverse propagator $P^{-1}_{AB}(\mathbf{p})$, there exists
an orthogonal matrix $U(\mathbf{p};m_0^2,p)$ such that
$U^TP^{-1}U$ is diagonal. If 
$v_A(\mathbf{p};m_0^2,p) \equiv U_{A1}(\mathbf{p};m_0^2,p)$
is the eigenvector that correspond to the zero eigenvalue for $\mathbf{p} = 0$ 
at the critical
point and $Q_{Aa}(\mathbf{p};m_0^2,p) \equiv U_{A,a+1}(\mathbf{p};m_0^2,p)$,
$a = 1,\ldots 4$ are the other eigenvectors, we define new fields
$\Phi_A(\mathbf{p})$ by writing
\begin{equation}
\Psi_A(\mathbf{p}) \equiv  \sum_B U_{AB} (\mathbf{p}) \Phi_B(\mathbf{p}) =
  v_A(\mathbf{p}) \phi(\mathbf{p}) + \sum_a Q_{Aa} (\mathbf{p}) 
  \varphi_a(\mathbf{p}),
\label{depPhi}
\end{equation}
where $\Phi = (\phi,\varphi_a)$. Eq.~\reff{depPhi} defines the fields 
$\Phi_A$ up to a sign. For definiteness we shall assume $v_A(\mathbf{p})$ 
to be such that, at the critical point, 
\be
   v_A(\mathbf{0}) = z_A/(\sum_B z_B^2)^{1/2}.
\ee
We do not specify the sign of $\varphi_a$ since it will 
not play any role in the following.

The new field $\phi$ corresponds to the zero mode, while the four fields
$\varphi_a$ are the noncritical (massive) modes. 
The effective Hamiltonian for the fields $\Phi$ has an expansion analogous 
to that presented in Eq.~\reff{effexp} for $\Psi$. The propagator 
$\hat{P}_{AB}(\mathbf{p})$ of $\Phi$ is 
\begin{eqnarray}
&& \hat{P}_{AB}(\mathbf{p}) = \sum_{CD} {P}_{CD}(\mathbf{p})
      U_{CA}(\mathbf{p}) U_{DB}(\mathbf{p}) ,
\end{eqnarray}
while the effective vertices are related to the previous ones by
\be 
 \hat{V}_{A_i,\ldots,A_n}^{(n)}(\mathbf{p}_1,\ldots,\mathbf{p}_n) = 
\sum_{B_1,\ldots,B_n}    
  {V}_{B_1,\ldots,B_n}^{(n)}(\mathbf{p}_1,\ldots,\mathbf{p}_n)
  U_{B_1A_1}(\mathbf{p}_1;m_0,p) \cdots
  U_{B_nA_n}(\mathbf{p}_n;m_0,p).
\ee
Note that $\hat{P}_{AB}(\mathbf{p})$ is diagonal by definition, i.e., 
$\hat{P}_{AB}(\mathbf{p}) = \delta_{AB} \hat{P}_{AA}(\mathbf{p})$.
Relation (\ref{detPinv-p0}) implies 
\be
\hat{P}_{11}(\mathbf{0}) \sim s_1 \sim (p-p_c),(m_0^2 - m_{0c}^2)^2
\ee
close to the critical point.
In Appendix \ref{AppA} 
we prove  a very general set of identities among the 
effective vertices.  In particular, we show that at the critical point
we also have
\be
\hat{V}_{111}(\mathbf{0},\mathbf{0},\mathbf{0})  = 0.
\ee
This result will play an important role in the following.

\section{Effective Hamiltonian for the zero mode}  \label{sec5}
\label{effHam-zeromode}

As we have discussed in the previous Section, the propagator $P$ has 
a zero mode at the critical point. Thus, in a neighbourhood of the 
critical point the standard $1/N$ expansion breaks down and a much more careful 
treatment of the zero mode is needed. For this purpose, we are now going to 
integrate out the massive modes, obtaining an effective Hamiltonian for the 
field $\phi$. More precisely, we define
\be
e^{-{\cal H}_{\rm eff}[\phi]} = 
   \int \prod_{xa} d\varphi_{xa} e^{-{\cal H}[\Phi]}.
\ee
The effective Hamiltonian ${\cal H}_{\rm eff}[\phi]$ 
has an expansion of the form
\begin{eqnarray}
{\cal H}_{\rm eff}[\phi] &=& {1\over \sqrt{N}} \tilde{H} \phi(\mathbf{0}) + 
        {1\over2} \int_{\mathbf{p}} 
   \phi(-\mathbf{p}) \tilde{P}^{-1}(\mathbf{p}) \phi(\mathbf{p})
\label{effexp2} \\
  &+& \sum_{n=3} {1\over n!} {1\over N^{n/2-1} }
  \int_{\mathbf{p}_1} \cdots \int_{\mathbf{p}_n} 
\delta(\sum_i \mathbf{p}_i) \; 
   \tilde{V}^{(n)} (\mathbf{p}_1,\ldots,\mathbf{p}_n)
   \phi(\mathbf{p}_1)\cdots \phi(\mathbf{p}_n),
\nonumber 
\end{eqnarray}
where vertices and propagators also depend on $N$ and have an expansion 
of the form
\begin{eqnarray}
  && \tilde{H} = \sum_{m=0} {1\over N^m} \tilde{H}_m, \nonumber \\
  && \tilde{P}^{-1}(\mathbf{p}) = 
       \sum_{m=0} {1\over N^m} \tilde{P}_m^{-1}(\mathbf{p}), \nonumber \\
   && \tilde{V}^{(n)} (\mathbf{p}_1,\ldots,\mathbf{p}_n) = 
     \sum_{m=0} {1\over N^m} 
    \tilde{V}^{(n)}_m (\mathbf{p}_1,\ldots,\mathbf{p}_n).
\end{eqnarray}
We report here the explicit expressions that we shall need in the following:
\begin{eqnarray}
&& \tilde{H}_0 = {1\over2} \int_{\mathbf{p}} \sum_{ab}
         \hat{V}_{1ab}(\mathbf{0},\mathbf{p},-\mathbf{p}) 
                       \hat{P}_{ab}(\mathbf{p}), 
\\
&& \tilde{P}^{-1}_0 (\mathbf{p})= \hat{P}_{11}^{-1} (\mathbf{p}),
\\
&& \tilde{P}^{-1}_1(\mathbf{p}) = {1\over2} \int_{\mathbf{q}}
   \left[ \sum_{ab} 
    \hat{V}_{11ab}^{(4)}(\mathbf{p},-\mathbf{p},\mathbf{q},-\mathbf{q}) 
        \hat{P}_{ab}(\mathbf{q}) -
   \right. 
\nonumber \\
&& \qquad\qquad\qquad  
       \sum_{abcd} \hat{V}_{11a}^{(3)}(\mathbf{p},-\mathbf{p},\mathbf{0})
       \hat{P}_{ab}(\mathbf{0})
       \hat{V}_{bcd}^{(3)}(\mathbf{0},\mathbf{q},-\mathbf{q}) \
               \hat{P}_{cd}(\mathbf{q}) -
\nonumber \\
&& \qquad\qquad\qquad  \left.
       \sum_{abcd}
       \hat{V}_{1ab}^{(3)}(\mathbf{p},\mathbf{q},-\mathbf{p}-\mathbf{q})
       \hat{V}_{1cd}^{(3)}(\mathbf{p},\mathbf{q},-\mathbf{p}-\mathbf{q})
       \hat{P}_{ac}(\mathbf{q}) \hat{P}_{bd}(\mathbf{p}+\mathbf{q})\right],
\\
&& \tilde{V}^{(3)}_0(\mathbf{p},\mathbf{q},\mathbf{r}) = 
 \hat{V}^{(3)}_{111}(\mathbf{p},\mathbf{q},\mathbf{r}) ,
\\
&& \tilde{V}^{(4)}_0(\mathbf{p},\mathbf{q},\mathbf{r},\mathbf{s}) = 
 \hat{V}^{(4)}_{1111}(\mathbf{p},\mathbf{q},\mathbf{r},\mathbf{s}) -
\nonumber \\
&& \qquad
   \sum_{ab} \hat{V}_{11a}^{(3)}(\mathbf{p},\mathbf{q},-\mathbf{p}-\mathbf{q})
          \hat{V}_{11b}^{(3)}(\mathbf{r},\mathbf{s},-\mathbf{r}-\mathbf{s})
          \hat{P}_{ab}(\mathbf{p}+\mathbf{q})
  + \hbox{\rm two permutations},
\\
&& \tilde{V}^{(5)}_0(\mathbf{p},\mathbf{q},\mathbf{r},\mathbf{s},\mathbf{t}) = 
 \hat{V}^{(5)}_{1111}(\mathbf{p},\mathbf{q},\mathbf{r},\mathbf{s},\mathbf{t}) -
\nonumber \\
&& \qquad \left[
   \sum_{ab} \hat{V}^{(3)}_{11a}(\mathbf{p},\mathbf{q},-\mathbf{p}-\mathbf{q})
  \hat{V}^{(4)}_{111b}(\mathbf{r},\mathbf{s},\mathbf{t},
         -\mathbf{r}-\mathbf{s}-\mathbf{t})
          \hat{P}_{ab}(\mathbf{p}+\mathbf{q}) + 
      \hbox{\rm 9 permutations}\right] + 
\nonumber \\
&& \qquad \left[
   \sum_{abcd} \hat{V}^{(3)}_{11a}(\mathbf{p},\mathbf{q},-\mathbf{p}-\mathbf{q})
  \hat{V}^{(3)}_{1bc}(\mathbf{r},\mathbf{p}+\mathbf{q},\mathbf{s}+\mathbf{t})
  \hat{V}^{(3)}_{11d}(\mathbf{s},\mathbf{t},-\mathbf{s}-\mathbf{t})
  \times \right. 
\nonumber \\
&&  \qquad\qquad \left. \times 
          \hat{P}_{ab}(\mathbf{p}+\mathbf{q}) 
          \hat{P}_{cd}(\mathbf{s}+\mathbf{t}) 
       + 
      \hbox{\rm 14 permutations}\right],
\end{eqnarray}
where $a$, $b$, $c$, $d$ run from 1 to 4 over the massive modes. 

The identities presented in App.~\ref{AppA} allow us to derive several 
relations among the effective vertices at the critical point. We have 
for $p = p_c$ and $m_0^2 = m_{0c}^2$
\begin{eqnarray}
&& \tilde{P}^{-1}_0 (\mathbf{0}) = 
  {\partial \tilde{P}^{-1}_0 (\mathbf{0})\over \partial m_0^2} = 0,
\label{id-P-CP}
\\
&& \tilde{V}_0^{(3)} (\mathbf{0},\mathbf{0},\mathbf{0}) = 0, 
\label{id-V3-CP}
\\
&& \left[{\partial\over \partial m_0^2}\tilde{V}_0^{(3)} 
      (\mathbf{0},\mathbf{0},\mathbf{0})\right]^2 = 
  {\partial^2 \tilde{P}^{-1}_0 (\mathbf{0})\over \partial (m_0^2)^2} 
 \tilde{V}_0^{(4)} (\mathbf{0},\mathbf{0},\mathbf{0},\mathbf{0}) .
\label{id-V4-CP}
\end{eqnarray}

Relation \reff{id-P-CP} clearly shows that the standard $1/N$ expansion 
fails close to the critical point. Indeed, outside the critical point, 
$\tilde{P}^{-1}_0 (\mathbf{p})$ is nonsingular and for large $N$ it is 
enough to 
expand the interaction Hamiltonian in powers of $1/N$. On the other hand, this
not possible at the critical point. Since also the three-leg vertex vanishes
at zero momentum in this case, cf.~Eq.~\reff{id-V3-CP}, the zero-momentum
leading term is the quartic one. 
Since the coupling constant is proportional 
to $1/N$, the model effectively corresponds to a weakly
coupled $\phi^4$ theory. 
In order to have a stable $\phi^4$ theory, we must also have 
$\tilde{V}_0^{(4)} (\mathbf{0},\mathbf{0},\mathbf{0},\mathbf{0}) > 0$.
For a generic solution of the gap equations satisfying 
Eq.~\reff{eq-punto-critico}, this is not {\em a priori} guaranteed.
Note, however, that if 
$\tilde{V}_0^{(4)} (\mathbf{0},\mathbf{0},\mathbf{0},\mathbf{0}) < 0$,
then, for $p=p_c$, we have 
$\tilde{P}^{-1}_0(\mathbf{0}) \approx a (m_0^2 - m_{0c}^2)^2$,
with $a < 0$, as a consequence of Eq.~\reff{id-V4-CP}: the propagator
has a {\em negative} mass for $N=\infty$. We believe---but we have 
not been able to prove---that such a phenomenon signals the fact that
the solution we are considering is not the relevant one. We expect 
the existence of another solution of the gap equation 
\reff{puntosella2} with a lower free energy.

The weakly
coupled $\phi^4$ theory shows an interesting crossover limit. If 
one neglects fluctuations it corresponds to tune $p$ and $m_0^2$ so that 
$\tilde{H}$ and $\tilde{P}^{-1}(\mathbf{0})$ go to zero as $N\to \infty$,
in such a way that $\tilde{H} N$ and $\tilde{P}^{-1}(\mathbf{0}) N$
remain constant. In this limit, Ising behavior is observed when 
the two scaling variables go to zero, while mean-field behavior is 
observed in the opposite case. Fluctuations change the simple scaling 
forms reported above and one must consider two additive renormalizations. 
A complete discussion will be  reported below in Sec.~\ref{sec6}. 

Finally, we wish to change 
again the definition of the field so that the effective zero-momentum 
three-leg vertex vanishes for all $p$ and $m_0^2$ in the limit $N\to \infty$.
For this purpose we now define a new field
\be
\alpha \chi(\mathbf{p}) = \phi(\mathbf{p}) + k \delta(\mathbf{p}),
\ee
where $\alpha$ and $k$ are functions of $p$ and $m_0^2$ to be fixed. 
The function $k$ is fixed by requiring that the 
large-$N$ zero-momentum three-leg vertex vanishes. 
Apparently, all $\tilde{V}^{(n)}$ contribute in this calculation. 
However, because of Eq.~\reff{id-V3-CP}, 
$\tilde{V}^{(3)}_0 (\mathbf{0},\mathbf{0},\mathbf{0}) $
vanishes at the critical point. As we already mentioned the 
interesting limit corresponds to considering 
$\Delta_m\equiv m_0^2 - m_{0c}^2\to 0$ and 
$\Delta_p\equiv p - p_c  \to 0$ together with $N\to \infty$. We will 
show in Sec.~\ref{sec7} that this limit should be 
taken keeping fixed $\Delta_m N^{1/3}$ and $\Delta_p N$, so that 
$\tilde{V}^{(3)}_0 (\mathbf{0},\mathbf{0},\mathbf{0}) $ is effectively of 
order $N^{-1/3}$. Therefore, the equation defining $k$ 
can be written in a compact form as 
\begin{equation}
{a_3 \over N^{5/6}} + 
  \sum_{n\ge 4} {a_n k^{n-3}\over N^{n/2 - 1}} = 0,
\label{eqaa}
\end{equation}
where $a_n \approx a_{n0} + a_{n1}/N$ is the contribution of the 
$n$-point vertex at zero momentum. Eq.~(\ref{eqaa}) can be rewritten as 
\be
\sum_{n=1}^\infty {a_{n+3}\over N^{(n-1)/3}}
        \left( {k\over N^{1/6}}\right)^n = - a_3 
\ee
which shows that $k$ has an expansion of the form
\be 
k = k_0 N^{1/6} [1 + k_1 N^{-1/3} + O(N^{-2/3})].
\ee
The leading constant $k_0$ depends only on the three- and four-leg 
vertex, the constant $k_1$ depends also on the five-leg vertex, and 
so on. The explicit calculation gives
\begin{eqnarray}
k &=& \sqrt{N} {\tilde{V}^{(3)}_0 (\mathbf{0},\mathbf{0},\mathbf{0}) \over 
       \tilde{V}^{(4)}_0 (\mathbf{0},\mathbf{0},\mathbf{0},\mathbf{0}) } + 
\nonumber \\
&& +    {\sqrt{N}\over2} 
  {\tilde{V}^{(3)}_0 (\mathbf{0},\mathbf{0},\mathbf{0})^2 
\tilde{V}^{(5)}_0 (\mathbf{0},\mathbf{0},\mathbf{0},\mathbf{0},\mathbf{0})
     \over
  [\tilde{V}^{(4)}_0 (\mathbf{0},\mathbf{0},\mathbf{0},\mathbf{0})]^3 } + 
    O(N^{-1/2}).
\end{eqnarray}
By performing this rescaling, the
$k$-leg $\phi$-vertex that scales with $N$ as $N^{1 - k/2}$ 
(except for $k\le 3$) 
gives a contribution to the $m$-leg $\chi$ vertex (of course $m\le k$) of order 
$N^{1 - m/2 + (m-k)/3}$, which is therefore always subleading. 
Taking this result into account we obtain 
\begin{eqnarray}
{\cal H}_{\rm eff} &=& H \chi(\mathbf{0}) + 
     {1\over 2} \int_{\mathbf{p}} \chi(\mathbf{p}) \chi(-\mathbf{p}) 
        \bar{P}^{-1} (\mathbf{p}) + 
\nonumber \\
&& +  {1\over 3! \sqrt{N}} \int_{\mathbf{p},\mathbf{q}}
      \bar{V}^{(3)}(\mathbf{p},\mathbf{q},-\mathbf{p}-\mathbf{q})
      \chi(\mathbf{p}) \chi(\mathbf{q}) \chi(-\mathbf{p}-\mathbf{q})
\nonumber \\
  &&  + {1\over 4! N} \int_{\mathbf{p},\mathbf{q},\mathbf{r}}
      \bar{V}^{(4)}(\mathbf{p},\mathbf{q},\mathbf{r},
           -\mathbf{p}-\mathbf{q}-\mathbf{r})
      \chi(\mathbf{p}) \chi(\mathbf{q}) \chi(\mathbf{r})
      \chi(-\mathbf{p}-\mathbf{q}-\mathbf{r}) + \ldots
\label{caleffchi}
\end{eqnarray}
where 
\begin{eqnarray} 
&& H={\alpha \tilde{H}_0\over \sqrt{N}} - 
    \alpha k \left[\tilde{P}_{0}^{-1}(\mathbf{0}) + 
                  {1\over N} \tilde{P}_{1}^{-1}(\mathbf{0})\right]+
    {\alpha k^2\over 2\sqrt{N}} \tilde{V}^{(3)}_{0}
         (\mathbf{0},\mathbf{0},\mathbf{0}) - 
    {\alpha k^3\over 6N} \tilde{V}^{(4)}_{0}
         (\mathbf{0},\mathbf{0},\mathbf{0},\mathbf{0})  
\nonumber \\
&& \qquad\qquad 
    + {\alpha k^4\over 24{N}^{3/2}} \tilde{V}^{(5)}_{0}
         (\mathbf{0},\mathbf{0},\mathbf{0},\mathbf{0},\mathbf{0}) 
+ O(N^{-7/6}),
\nonumber \\
&& \bar{P}^{-1} (\mathbf{p}) = 
   \alpha^2 \tilde{P}_{0}^{-1}(\mathbf{p}) + 
   {\alpha^2\over N} \tilde{P}_{1}^{-1}(\mathbf{p}) -
   {\alpha^2 k\over \sqrt{N}} \tilde{V}^{(3)}_0
         (\mathbf{0},\mathbf{p},-\mathbf{p}) + 
   {\alpha^2 k^2\over 2 N} 
     \tilde{V}^{(4)}_{0}(\mathbf{0},\mathbf{0},\mathbf{p},-\mathbf{p}) 
\nonumber \\ 
 && \qquad\qquad
   - {\alpha^2 k^3\over 6 N^{3/2}}  
     \tilde{V}^{(5)}_{0}(\mathbf{0},\mathbf{0},\mathbf{0},
               \mathbf{p},-\mathbf{p}) + O(N^{-4/3}), 
\nonumber \\ [3mm]
&& \bar{V}^{(3)}(\mathbf{p},\mathbf{q},\mathbf{r}) = 
   \alpha^3 \tilde{V}^{(3)}_{0}(\mathbf{p},\mathbf{q},\mathbf{r}) - 
   {\alpha^3 k\over \sqrt{N}} 
     \tilde{V}^{(4)}_{0}(\mathbf{0},\mathbf{p},\mathbf{q},\mathbf{r}) + 
     O(N^{-2/3}), 
\nonumber \\ [3mm]
&& \bar{V}^{(4)}(\mathbf{p},\mathbf{q},\mathbf{r},\mathbf{s}) =
   \alpha^4 \tilde{V}^{(4)}_0(\mathbf{p},\mathbf{q},\mathbf{r},\mathbf{s})
     + O(N^{-1/3}), 
\end{eqnarray} 
where $\alpha = \alpha_0 + {\alpha_1 N^{-2/3}} + O(N^{-4/3})$ 
is fixed by requiring that 
\begin{equation}
\bar{P}^{-1}(\mathbf{p}) - \bar{P}^{-1}(\mathbf{0}) \equiv 
K(\mathbf{p}) \approx  \mathbf{p}^2,
\end{equation}
for $\mathbf{p} \to 0$.

\section{Intermezzo: The critical crossover limit} \label{sec6}

In this section we present a general discussion of the critical
behavior of a generic weakly coupled lattice $\phi^4$ theory, as 
is the effective Hamiltonian \reff{caleffchi} for the zero mode.
In particular, we show the existence of a critical limit, the critical
crossover limit, that allows one to interpolate between the 
Ising behavior and the mean-field behavior. The arguments presented 
here closely follow the discussion of Ref.~\cite{PRV-99} for medium-range 
models. The main difference is the appearance of odd $\phi^3$ interactions
that require additional renormalizations.

\subsection{General considerations}

We wish now to discuss the critical behavior of a model with 
Hamiltonian
\begin{eqnarray}
{\cal H}_{\rm eff} &=& H \varphi(\mathbf{0}) + 
      {1\over 2} \int_\mathbf{p} [K(\mathbf{p}) + r] 
      \varphi(\mathbf{p}) \varphi(-\mathbf{p}) 
\nonumber \\ 
&& +  {\sqrt{u}\over 3!} \int_\mathbf{p} \int_\mathbf{q}
                V^{(3)}(\mathbf{p},\mathbf{q},-\mathbf{p}-\mathbf{q}) 
                \varphi(\mathbf{p}) \varphi(\mathbf{q}) 
                \varphi(-\mathbf{p}-\mathbf{q}) 
\nonumber \\ 
&& + 
      {u\over4!} \int_\mathbf{p} \int_\mathbf{q} \int_\mathbf{s}
         V^{(4)}(\mathbf{p},\mathbf{q},\mathbf{s},
                 -\mathbf{p}-\mathbf{q}-\mathbf{s}) 
           \varphi(\mathbf{p}) \varphi(\mathbf{q}) 
           \varphi(\mathbf{s}) \varphi(-\mathbf{p}-\mathbf{q}-\mathbf{s})
\label{Heff-ccl}
\end{eqnarray}
on a square lattice. Here the integrations are extended over the first 
Brillouin zone, $\varphi(\mathbf{p})$ is the Fourier transform of the 
fundamental field which is normalized so that 
$K(\mathbf{p}) \approx \mathbf{p}^2$ for $\mathbf{p}\to 0$. 
By properly normalizing $u$ (which is assumed to be positive) 
we can also require $V^{(4)}(\mathbf{0},\mathbf{0},\mathbf{0},\mathbf{0}) = 1$.
We shall only consider the case in which the three-leg vertex vanishes 
at zero external momenta, i.e. $V^{(3)}(\mathbf{0},\mathbf{0},\mathbf{0}) = 0$. 
This is an important simplifying assumption. Indeed, this means that the cubic 
interaction has the form $\varphi^2 \Box \varphi$, which is 
a renormalization-group irrelevant term according to power counting. 

If $V^{(3)}(\mathbf{p},\mathbf{q},\mathbf{r}) = 0$ for all momenta, 
there is an interesting critical
limit, the so-called critical crossover limit 
\cite{BB-85}. Indeed, one can show that,
by properly defining a function $r_c(u)$, in the limit
$u\to 0$, $t \equiv r - r_c(u)\to 0$, and $H\to 0$ at fixed $H/u$ and $t/u$,
the $n$-point susceptibility $\chi_n$, i.e. the zero-momentum 
$n$-point connected correlation function,  has the scaling form
\begin{equation}
  \chi_{n} \approx u t^{-n} f_n^{\rm symm} (t/u,H/u).
\label{scaling-cross-lim}
\end{equation}
The scaling function $f_n^{\rm symm}(x,y)$ is universal, i.e. it does 
not depend on the explicit form of $K(\mathbf{p})$ and of 
$V^{(4)}(\mathbf{p},\mathbf{q},\mathbf{r},\mathbf{s})$
as long as the normalization conditions fixed above are satisfied.
The definition of $r_c(u)$ has been discussed in Ref.~\cite{BB-85} for the 
continuum model and in Ref.~\cite{PRV-99} for the more complex case of 
medium-range models. It is of order $u$ and is obtained by requiring 
$\chi_2$ to scale
according to Eq.~(\ref{scaling-cross-lim}) in the critical crossover limit.
If $r_c(u) = r_1 u + O(u^2)$, then at one loop we have
\begin{equation}
\chi_2 = {1\over t} - {u\over t^2} \left[r_1 + {1\over2} 
       \int_\mathbf{p} 
   {V^{(4)}(\mathbf{0},\mathbf{0},\mathbf{p},-\mathbf{p})\over 
       K(\mathbf{p}) + t}\right] + 
       O(u^2).
\end{equation}
For $t\to 0$
\begin{equation}
\int_\mathbf{p}  {V^{(4)}(\mathbf{0} ,\mathbf{0} ,\mathbf{p},-\mathbf{p})\over 
   K(\mathbf{p}) + t} \approx
  \int_\mathbf{p} {1\over \hat{p}^2 + t} + {\rm constant}\, 
  = - {1\over 4\pi} \ln t + {\rm constant},
\end{equation}
where $\hat{p}^2 = 4 \sin^2(p_1/2) + 4 \sin^2(p_2/2)$. Thus, we obtain 
\begin{equation} 
\chi_2 = {1\over t} - {u\over t^2}\left[
    r_1 - {1\over 8\pi} \ln u - {1\over 8\pi} \ln {t\over u} + 
    {\rm constant}\, \right] + O(u^2).  
\end{equation} 
Therefore, if we define
\begin{equation}
r_c(u) = {u\over 8\pi} \ln u,
\label{rcu-phi4}
\end{equation}
the susceptibility $\chi_2$ scales according to Eq.~(\ref{scaling-cross-lim}).
No other correction is needed when considering the higher-order
contributions and thus Eq.~(\ref{rcu-phi4}) gives an exact nonperturbative 
definition of $r_c(u)$. Such a result can be easily understood.
In the continuum two-dimensional theory there is only one primitively 
divergent graph, the one-loop tadpole, and therefore only a one-loop 
mass counterterm is needed to make the theory finite. The same holds for the 
lattice model in the critical crossover limit, which is the limit in
which the lattice theory goes over to the continuum model.

Now, let us consider the contributions of the three-leg vertex. 
Since this is a renormalization-group irrelevant term,
the only changes we expect concern the renormalization of the bare 
parameters. Indeed, we will now show 
that, if one properly defines functions $r_c(u)$ and $h_c(u)$, then
in the limit 
$u\to 0$, $t \equiv r - r_c(u)\to 0$, and $h \equiv H - h_c(u)\to 0$ 
at fixed $h/u$ and $t/u$,
the $n$-point susceptibility $\chi_n$ has the scaling form
(\ref{scaling-cross-lim}) with $h$ replacing $H$ and 
with the same scaling functions of the symmetric case.

We will first prove this result at two loops and then we will give a general
argument that applies to all perturbative orders.
Notice that it is enough to consider the case $h = 0$. Indeed, if 
Eq.~(\ref{scaling-cross-lim}) is valid for $h=0$ and any $n$, then
\begin{eqnarray}
\chi_n(h) &=& \sum_{m=0} {1\over m!} \chi_{n+m}(h=0) h^m \approx 
            \sum_{m=0} {1\over m!} u t^{-n-m} f_{n+m}^{\rm symm}(t/u,0) h^m = 
\nonumber \\
          &=& u t^{-n} \sum_{m=0} 
           {1\over m!} f_{n+m}^{\rm symm}(t/u,0) (t/u)^{-m} (h/u)^m.
\end{eqnarray}
which proves Eq.~(\ref{scaling-cross-lim}) for all values of $h$.

\begin{figure}[t]
\centerline{\epsfig{height=10truecm,file=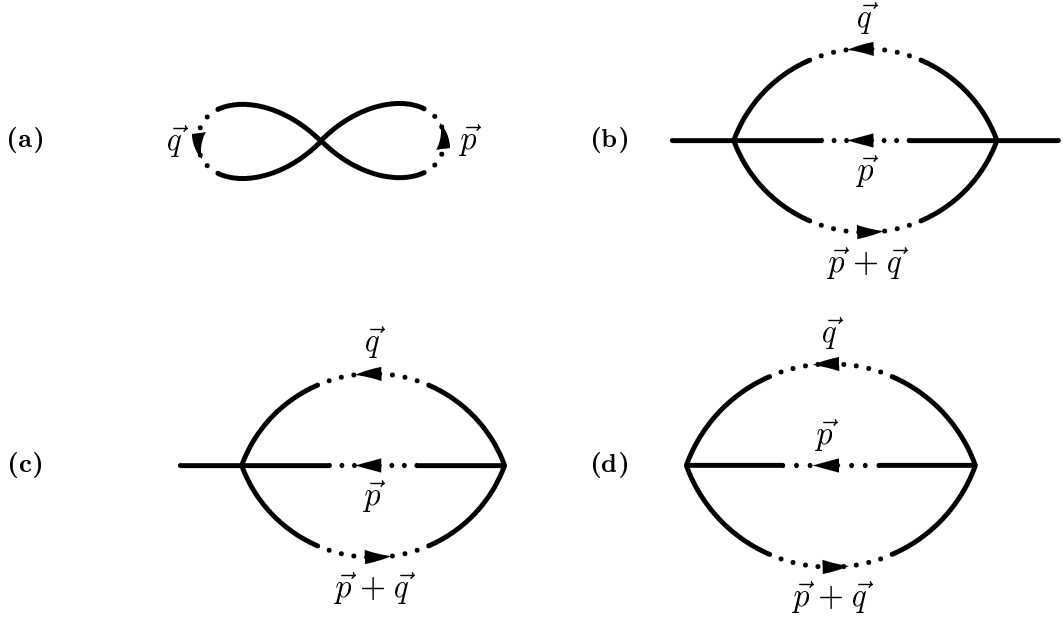}}
\caption{The four topologies appearing at two loops. Dots indicate 
parts of the graphs with additional external legs. }
\label{figgraph}
\end{figure}

\subsection{Explicit two-loop perturbative calculation} \label{twoloop-ccl}

At tree level, all contributions to $\chi_n$ that contain three-leg vertices
vanish because $V^{(3)}(\mathbf{0},\mathbf{0},\mathbf{0}) = 0$. 
Therefore, $\chi_n$ is identical to 
$\chi_n$ in the $\mathbb{Z}_2$-symmetric 
$\varphi^4$ theory; in particular $\chi_{2n+1} = 0$. 
At one-loop order, graphs contributing to $\chi_n$ are formed by a single loop
made of $a$ three-leg vertices and of $b$ four-leg vertices, 
with $a + 2 b = n$. Each of them contributes a term of the form
\begin{equation}
{u^{n/2}\over t^n} 
   \int_\mathbf{p} {V_3^a(\mathbf{p}) V_4^b(\mathbf{p})\over 
                   [K(\mathbf{p}) + t]^{a+b}},
\end{equation}
where $V_3(\mathbf{p}) \equiv V^{(3)}(\mathbf{0},\mathbf{p},-\mathbf{p})$ and 
$V_4(\mathbf{p}) \equiv V^{(4)}(\mathbf{0},\mathbf{0},\mathbf{p},-\mathbf{p})$.
The leading contribution for $t\to 0$ is obtained by replacing 
each quantity with its small-$\mathbf{p}$ behavior, i.e. 
$V_4(\mathbf{p})\approx 1$, 
$V_3(\mathbf{p}) \sim \hat{p}^2$, and $K(\mathbf{p}) \approx \hat{p}^2$,
with $\hat{p}^2 = 4 \sin^2(p_1/2) + 4 \sin^2(p_2/2)$. Then, we obtain 
a contribution proportional to 
\begin{equation}
{u^{n/2}\over t^n} \int_\mathbf{p}
     {(\hat{p}^2)^a \over (\hat{p}^2 + t)^{a+b}}. 
\end{equation} 
Now, for $t\to 0$ we have 
\begin{equation} 
\int_\mathbf{p}
     {(\hat{p}^2)^a \over (\hat{p}^2 + t)^{a+b}}
   \sim \cases{1              & for $b=0$ \cr
            \ln t         & for $b=1$ \cr
            t^{1-b}        & for $b\ge 2$.
           }
\label{int-1loop}
\end{equation}
Therefore, for $t\to 0$, $u\to 0$ at fixed $t/u$, we have 
\begin{equation}  
t^n \chi_n/u \sim \cases{u^{n/2-1} & for $b=0$ \cr  
                         u^{n/2-1} \ln t & for $b=1$ \cr
                         u^{n/2-b} = u^{a/2} & for $b\ge 2$.
           }
\end{equation}
Thus, all contributions vanish except those with: (i) $n=1$, $a=1$, $b=0$;
(ii) $n=2$, $a=2$, $b=0$; 
(iii) $n$ even, $b=n/2$, $a=0$. Contributions (iii) are those 
that appear in the standard theory without $\varphi^3$ interaction. 
Let us now show that contributions (i) and (ii) can be eliminated 
by redefining $r_c(u)$ and $h_c(u)$. Consider first $\chi_1$. At one loop 
we have 
\begin{equation}
{t\over u} \chi_1 = - {h_c(u)\over u} - {1\over 2 \sqrt{u}} 
\int_\mathbf{p} {V_3(\mathbf{p})\over K(\mathbf{p}) + t} + O (\sqrt{u}).
\label{chi1-oneloop}
\end{equation}
For $t\to 0$ we have 
\begin{equation}
\int_\mathbf{p} {V_3(\mathbf{p})\over K(\mathbf{p}) + t} = 
\int_\mathbf{p} {V_3(\mathbf{p})\over K(\mathbf{p})} + O(t\ln t).
\end{equation}
Thus, if we define 
\begin{equation}
h_c(u) = - {1\over2} \sqrt{u} \int_\mathbf{p} {V_3(\mathbf{p})\over 
        K(\mathbf{p})}, 
\label{def-hc-App}
\end{equation}
then 
$t\chi_1/u \sim O(t \ln t/\sqrt{u},\sqrt{u}) \to 0$ in the critical crossover 
limit.

Now, let us consider the two-point function. At one loop we have 
\begin{eqnarray}
{t^2\over u}\chi_2 \approx {t\over u} - {r_c(u)\over u} - 
 {1\over 2} \int_\mathbf{p} \left[
   {V_4(\mathbf{p})\over  K(\mathbf{p}) + t} - 
   {V_3(\mathbf{p})^2 \over (K(\mathbf{p}) + t)^2} \right].
\label{chi2-oneloop}
\end{eqnarray}
The first one-loop term is the contribution of the tadpole that has to be 
considered in the pure $\varphi^4$ theory and which requires an appropriate 
subtraction to scale correctly, cf. Eq.~(\ref{rcu-phi4}).
The second one is due to the 
three-leg vertex and is finite for $t\to 0$. Therefore, if we define 
\begin{equation}
r_c(u) = r_c^{\rm symm}(u) + 
     {u\over 2} \int_\mathbf{p}
     {V_3(\mathbf{p})^2 \over K(\mathbf{p})^2},
\label{def-rc-App}
\end{equation}
where $r_c^{\rm symm}(u)$ is given by Eq.~(\ref{rcu-phi4}), we cancel all 
contributions of the $\varphi^3$ vertex. 

Now, let us repeat the same discussion at two loops, 
in order to understand the 
general mechanism. There are four different topologies that must be considered,
see Fig. \ref{figgraph}.

{\em Topology (a).}

A contribution to $\chi_n$ is proportional to 
\begin{equation}
{u^{n/2+1}\over t^n} 
  \int_\mathbf{p} \int_\mathbf{q} 
   V^{(4)}(\mathbf{p},-\mathbf{p},\mathbf{q},-\mathbf{q}) 
  {V_4(\mathbf{p})^a V_4(\mathbf{q})^b V_3(\mathbf{p})^c V_3(\mathbf{q})^d
      \over (K(\mathbf{p})+t)^{a+c+1} (K(\mathbf{q})+t)^{b+d+1}},
\end{equation}
where $2a+2b+c+d=n$. The leading contribution for $t\to 0$ 
is obtained by setting
$V^{(4)}(\mathbf{p},-\mathbf{p},\mathbf{q},-\mathbf{q}) \approx 
 V^{(4)}(\mathbf{0},\mathbf{0},\mathbf{0},\mathbf{0}) = 1$. Then,
the integral factorizes and
we can use Eq.~(\ref{int-1loop}). Ignoring logarithmic terms, we see that 
the corresponding contribution to $t^n \chi_n/u$ scales as 
\begin{equation}
  \left({u\over t}\right)^{n/2} t^{n/2-a-b} = 
  \left({u\over t}\right)^{n/2} t^{(c+d)/2}\; .
\end{equation}
Thus, a nonvanishing contribution is obtained only for $c+d=0$. Three-leg 
vertices can be neglected in the critical crossover limit.

{\em Topology (b).}

A contribution to $\chi_n$ has the form
\begin{eqnarray}
&& {u^{n/2+1}\over t^n}
  \int_\mathbf{p} \int_\mathbf{q} 
   [V^{(4)}(\mathbf{0},\mathbf{p},\mathbf{q},-\mathbf{p}-\mathbf{q})]^2
\nonumber \\
  && \qquad \times 
    {V_4(\mathbf{p})^a V_4(\mathbf{q})^b V_4(\mathbf{p}+\mathbf{q})^c 
     V_3(\mathbf{p})^d V_3(\mathbf{q})^e V_3(\mathbf{p}+\mathbf{q})^f \over 
   (K(\mathbf{p})+t)^{a+d+1} (K(\mathbf{q})+t)^{b+e+1} 
   (K(\mathbf{p}+\mathbf{q})+t)^{c+f+1}},
\end{eqnarray}
where $2(a+b+c)+d+e+f=n-2$.
The leading infrared contribution is obtained by approximating all 
expressions with their small-$p$ behavior. Therefore, we can write for $t\to 0$
\begin{equation}
{u^{n/2+1}\over t^n}
  \int_\mathbf{p} \int_\mathbf{q}
  {(\hat{p}^2)^d (\hat{q}^2)^e (\widehat{p+q}^2)^f \over
   (\hat{p}^2+t)^{a+d+1} (\hat{q}^2+t)^{b+e+1} (\widehat{p+q}^2+t)^{c+f+1}}.
\end{equation} 
Integrals of this type can be easily evaluated. By writing 
$\hat{p}^2 = (\hat{p}^2 + t) - t$ in the numerator, we obtain integrals
\begin{equation} 
I_{mnr} \equiv  \int_\mathbf{p} \int_\mathbf{q}
  {1\over 
  (\hat{p}^2+t)^{m} (\hat{q}^2+t)^{n} (\widehat{p+q}^2+t)^{r}},
\label{int-mnr}
\end{equation}
with $m > 0$, $n > 0$, $r > 0$. Then,
we can rescale 
$\mathbf{p}\to t^{1/2} \mathbf{p}$, $\mathbf{q} \to t^{1/2} \mathbf{q}$ and 
extend the 
integration over all $\mathbb{R}^2$. This is possible since the corresponding 
continuum integral is finite. As a consequence the 
integral scales as $t^{2-m-n-r}$ and
the corresponding contribution to $t^n \chi_n/u$ scales as 
\begin{equation}
  \left({u\over t}\right)^{n/2} t^{n/2-a-b-c-1} = 
  \left({u\over t}\right)^{n/2} t^{(d+e+f)/2}.
\end{equation}
Thus, a nonvanishing contribution is obtained only for $d+e+f=0$. Three-leg 
vertices can be neglected in the critical crossover limit.

{\em Topology (c).}

A contribution to $\chi_n$ has the form
\begin{eqnarray}
&& {u^{n/2+1}\over t^n}
  \int_\mathbf{p} \int_\mathbf{q} 
   V^{(4)}(\mathbf{0},\mathbf{p},\mathbf{q},-\mathbf{p}-\mathbf{q})
    V^{(3)}(\mathbf{p},\mathbf{q},-\mathbf{p}-\mathbf{q})
\nonumber \\
&& \quad\qquad \times 
  {V_4(\mathbf{p})^a V_4(\mathbf{q})^b V_4(\mathbf{p}+\mathbf{q})^c 
   V_3(\mathbf{p})^d V_3(\mathbf{q})^e V_3(\mathbf{p}+\mathbf{q})^f \over 
   (K(\mathbf{p})+t)^{a+d+1} (K(\mathbf{q})+t)^{b+e+1} 
   (K(\mathbf{p}+\mathbf{q})+t)^{c+f+1}},
\label{topc}
\end{eqnarray}
where $2(a+b+c)+d+e+f=n-1$ and we assume without loss of generality 
$a\ge b \ge c$. We can now repeat the analysis performed for 
topology (b). We replace each quantity with its small-momentum behavior.
In particular, we can replace 
$V^{(3)}(\mathbf{p},\mathbf{q},-\mathbf{p}-\mathbf{q})$ with 
$(\hat{p}^2 + \hat{q}^2 + \widehat{p+q}^2)$. Then, 
we rewrite each contribution in terms of the integrals 
$I_{mnr}$, cf. Eq.~(\ref{int-mnr}). However, in this case it is possible that 
one (and only one) of the indices vanishes. If this the case, 
$I_{mnr}$ factorizes and we can use the one-loop result (\ref{int-1loop}).
A careful analysis shows 
that in all cases the integral scales as $t^{-a-b-c}$ for $t\to 0$ 
except when $b=c=0$. In this case, if $a\not= 0$ the integral scales as 
$t^{-a} \ln t$, while for $a=0$ it scales as $\log^2 t$.
Thus, ignoring logarithms all 
integrals scale as $t^{-a-b-c}$. Therefore, 
the corresponding contribution to $t^n \chi_n/u$ scales as 
\begin{equation}
  \left({u\over t}\right)^{n/2} t^{n/2-a-b-c} = 
  \left({u\over t}\right)^{n/2} t^{(d+e+f+1)/2}.
\end{equation}
These contributions always vanish. 

{\em Topology (d).}

A contribution to $\chi_n$ has the form
\begin{equation}
{u^{n/2+1}\over t^n}
  \int_\mathbf{p} \int_\mathbf{q} 
  [V^{(3)}(\mathbf{p},\mathbf{q},-\mathbf{p}-\mathbf{q})]^2
  {V_4(\mathbf{p})^a V_4(\mathbf{q})^b V_4(\mathbf{p}+\mathbf{q})^c 
   V_3(\mathbf{p})^d V_3(\mathbf{q})^e V_3(\mathbf{p}+\mathbf{q})^f \over 
   (K(\mathbf{p})+t)^{a+d+1} (K(\mathbf{q})+t)^{b+e+1} 
   (K(\mathbf{p}+\mathbf{q})+t)^{c+f+1}},
\end{equation}
where $2(a+b+c)+d+e+f=n$, $a\ge b\ge c$. 
We repeat the analysis done for topology (b) and (c). We find that the 
integral scales as 
\begin{equation}
t^{-a-b-c+1} \cases{\ln t/t   &  for $a=b=c=0$;  \cr
                    1/t        &  for $b=c=0$, $a \ge 1$; \cr
                    \ln t     &  for $c=0$, $b=1$, $a \ge 1$; \cr
                    {\rm constant }  &  otherwise.}
\end{equation}
Thus, if $b\not=0$ and $c\not=0$, ignoring logarithms, the contribution to 
$t^n\chi_n/u$ scales as 
\begin{equation}
\left({u\over t}\right)^{n/2} t^{n/2-a-b-c+1} = 
\left({u\over t}\right)^{n/2} t^{(d+e+f)/2+1} 
\end{equation}
These contributions therefore always vanish in the critical crossover limit. 
If, however, $b=c=0$, then 
\begin{equation}
\left({u\over t}\right)^{n/2} t^{n/2-a-b-c} =
\left({u\over t}\right)^{n/2} t^{(d+e+f)/2}
\end{equation} 
and thus one may have a finite (or a logarithmically divergent 
if $a = 0$) contribution for $d=e=f=0$. Let us now focus on this last
case in which $a=n/2 >0$ 
with $n$ even. Let us now show that these contributions are canceled by
the one-loop counterterm due to $r_c(u)$. Indeed, at two loops we obtain 
a contribution to $\chi_n$ of the form 
\begin{equation}
r_c(u) {u^{n/2}\over t^n} \int_\mathbf{p}
       {V^b_3(\mathbf{p}) V_4^a(\mathbf{p}) \over [K(\mathbf{p}) + t]^{a+b+1}} 
  = 
    r_c(u) {u^{n/2}\over t^n} \cases{\ln t  & for $b=0$ \cr 
                                     t^{-a} & for $b\ge 1$ 
        }
\end{equation} 
where $2 a + b =n$. Thus, contributions to $t^n\chi_n/u$ scale as 
$t^{n/2-a} = t^{b/2}$ and thus vanish unless $b=0$. 
Thus, for each even $n$ there are two contributions that should be 
considered: one with topology (d) and one associated with the counterterm
$r_c(u)$. Taking properly into account the combinatorial factors,
their sum is given by (of course, 
we only consider here the contribution to $r_c(u)$ due to the three-leg 
vertices)
\begin{equation} 
{u^{n/2+1}\over t^n} 
    \int_\mathbf{p} {1\over (K(\mathbf{p}) + t)^{a+1}} 
    \int_\mathbf{q} \left[
    {V^{(3)}(\mathbf{p},\mathbf{q},-\mathbf{p}-\mathbf{q})^2 \over 
      (K(\mathbf{q}) + t)(K(\mathbf{p}+\mathbf{q})+t)} - 
    {V^{(3)}(0,\mathbf{q},-\mathbf{q})^2 \over (K(\mathbf{q}) + t)^2}\right]
\end{equation}
For $a\ge 1$, the subtracted term improves the infrared behavior, and 
indeed the integral scales as $t^{-a+1} \ln t$  and is therefore 
irrelevant in the critical crossover limit. 

\subsection{Higher powers of the fields}

It is interesting to consider also Hamiltonians with 
higher powers of the field $\varphi$. 
The standard scaling argument indicates that these additional terms 
 are irrelevant for the 
critical behavior, but in principle they can contribute 
to the renormalization constants like $h_c$ and $r_c$. For this purpose let us 
suppose that the Hamiltonian contains also terms of the form 
\begin{equation}
\Delta{\cal H}_{\rm eff} = \sum_{k>4} {u^{(k-2)/2}\over k!}
    \int_{\mathbf{p}_1}\cdots \int_{\mathbf{p}_k}
\delta(\mathbf{p}_1 + \cdots \mathbf{p}_k) \, 
    V^{(k)}(\mathbf{p}_1,\cdots,\mathbf{p}_k) 
   \varphi(\mathbf{p}_1)\cdots \varphi(\mathbf{p}_k).
\end{equation}
Note the particular dependence on the coupling constant $u$, which is 
motivated by the large-$N$ calculation and is crucial in the argument 
reported below.

Let us consider the contributions of these additional terms. At tree level
we have an additional contribution to $\chi_n$ given by
\begin{equation}
\Delta\chi_n = {u^{(n-2)/2}\over t^n} V^{(n)}
    (\mathbf{0},\ldots,\mathbf{0})
\end{equation}
Thus, $t^n \Delta\chi_n/u \sim u^{(k-4)/2}$, that vanishes as $u\to 0$ for 
$k > 4$. At one-loop order there are additional contributions of the 
form 
\begin{equation}
   {u^{n/2}\over t^n} \int_\mathbf{p} (K(\mathbf{p}) + t)^{-\sum a_i} 
   V_3(\mathbf{p})^{a_3} \cdots V_{n+2}(\mathbf{p})^{a_{n+2}},
\end{equation}
where $V_k(\mathbf{p}) = 
   V^{(k)}(-\mathbf{p},\mathbf{p},\mathbf{0},\ldots,\mathbf{0})$ and 
$\sum_k(k-2)a_k = n$. 
Proceeding as in the previous Section, using Eq.~\reff{int-1loop}, and 
noting that $\sum_{k>4} a_k > 0$ by hypothesis, we obtain
\begin{equation}
t^n \Delta\chi_n/u \sim u^{n/2-1} t^{1 - \sum_{k > 3} a_k} \sim
   u^{{1\over2} a_3 + {1\over2} \sum_{k \ge 5} (k-4) a_k}.
\end{equation}
Here, possible logarithmic terms have been neglected. Then, since some 
$a_k$ with $k\ge 5$ is nonvanishing by hypothesis, we find that this
correction vanishes in the crossover limit. 
Therefore, no contribution survives at one loop. The same is expected at any
perturbative order.

\subsection{The general argument}

The discussion reported above shows that at two loops one can 
define a critical crossover limit with crossover functions that are identical 
to those of the symmetric theory. This is expected to be a general result
since formally the added interaction is irrelevant. 
This result can be understood diagrammatically. 

Consider the continuum theory with Hamiltonian
\begin{equation}
{\cal H} = {1\over2}\sum_\mu (\partial_\mu \phi)^2 + 
           {t\over2} \phi^2 + {\sqrt{u}\over 3!} \phi^2 \Box \phi + 
           {u\over 4!} \phi^4.
\end{equation}
Given an $l$-loop diagram contributing to the zero-momentum $n$-point 
irreducible correlation function, we can write the corresponding
Feynman integral $D$ as 
\begin{equation}
D \sim u^{N_4 + N_3/2} 
  \int_{\mathbf{p}_1} \cdots \int_{\mathbf{p}_l} \; 
    I(\mathbf{p}_1,\ldots,\mathbf{p}_l)
\label{scalingD}
\end{equation}
where $N_3$ and $N_4$ are the number of three-leg and four-leg vertices 
respectively. Using the topological relations $N_i = N_3 + N_4 + l - 1$ and 
$N_3 + 2 N_4 - 2l + 2 = n$, 
where $N_i$ is the number of internal lines, it is easy to see that 
\begin{equation} 
I(\lambda \mathbf{p}_1,\ldots,\lambda \mathbf{p}_l) = 
   \lambda^{2 (1 - N_4)} I(\mathbf{p}_1,\ldots,\mathbf{p}_l).
\label{scalingI}
\end{equation}
Thus, there are primitively divergent diagrams for any $n$: those with 
$N_4=0$ (and correspondingly $N_3 = n + 2l - 2$) are quadratically 
divergent, while those with $N_4 = 1$ (and $N_3 = n + 2l - 4$) are 
logarithmically divergent. Analogously, diagrams contributing to the 
second derivatives of the $n$-point irreducible correlation functions with
respect to the external momenta diverge logarithmically when $N_4 = 0$. 
Thus, the renormalized Hamiltonian contains an infinite number of 
counterterms and is given by
\begin{equation}
{\cal H}^{\rm ren} = {\cal H} + 
   \sum_n [Z_n(\Lambda) \phi^n + \zeta_n(\Lambda) \phi^{n-1}\Box\phi],
\end{equation}
where $\Lambda$ is a generic cutoff (in our explicit calculation 
$\Lambda$ is the inverse lattice spacing $1/a$, that never explicitly
appears in the calculation since we set $a=1$).
Now, let us show that all counterterms except those 
computed in the previous Section can be neglected in the critical 
crossover limit. Suppose that we wish to compute $Z_n(\Lambda)$ at 
$l$ loops. Keeping into account that the divergence may be quadratic or 
logarithmic, we expect the $l$-loop divergent contribution to $\chi_n$ to be 
of the form 
\begin{equation}
{u^{n/2+l-1}\over t^n} (a_1 \Lambda^2 + t P_1[\ln (\Lambda^2/t)] + 
      P_2[\ln (\Lambda^2/t)] ), 
\end{equation}
where $P_1(x)$ and $P_2(x)$ are polynomials and $a_1$ a constant.
Therefore, the contribution to $t^n\chi_n/u$ vanishes unless 
$n/2 + l - 2\le 0$. The only two cases satisfying this condition 
(of course $n\ge 1$ and $l\ge 1$) are $l=n=1$, $l=1$ and $n=2$, which are 
the cases considered before.
Let us now consider the contributions to $\bar{\chi}_{n,i}$, which is 
the first derivative of the $n$-point connected correlation function
with respect to the square of an external momentum $\mathbf{p}$ computed
at zero momentum. Since momenta scale as $t^{1/2}$, in the critical
crossover limit we should have 
$\bar{\chi}_{n,i}\approx u t^{-n-1} \bar{f}_n(t/u,H/u)$. 
The divergent contributions are logarithmic (diagrams with $N_4 = 0$) 
and therefore we expect an $l$-loop contribution of the form
\begin{equation}
{u^{n/2+l-1}\over t^n} (P[\ln (\Lambda^2/t)] + \hbox{\rm finite  terms}).
\end{equation}
Considering $t^{n+1} \bar{\chi}_{n,i}/u$, we see that this contribution 
always vanishes in the critical crossover limit. Therefore, 
the renormalization constants $\zeta_n(\Lambda)$ can be neglected. 
Thus, the only renormalizations needed are those that we have 
considered. Finally, let us show that correlation functions computed in 
the renormalized theory have the correct scaling behavior. Indeed, in
the renormalized theory diagrams scale canonically with possible logarithmic 
corrections. Therefore, $D$ defined in Eq.~(\ref{scalingD}) scales as 
$u^{N_4 + N_3/2} t^{1-N_4} \times$logs, so that the contribution of 
$D$ to $t^n \chi_n/u$ scales as 
\begin{equation}
   {t^n\over u} \times {1\over t^n} \times u^{N_4 + N_3/2} t^{1 - N_4} 
   \sim u^{N_3/2}.
\end{equation}
Therefore, the only nonvanishing diagrams have $N_3=0$, confirming the 
claim that three-leg vertices do not play any role.

\subsection{A unique definition for the renormalization functions
$h_c(u)$ and $r_c(u)$} \label{sec6.5}

In this Section we wish to discuss again the definition of $r_c(u)$
and $h_c(u)$. It is obvious that these functions are not uniquely defined,
since one can add a term proportional to $u$ without modifying the 
scaling behavior. We wish now to fix this ambiguity by 
requiring that $t = h = 0$ corresponds to the critical point. 

It is easy to see that no modifications are needed for $h_c(u)$. 
Indeed, with the choice (\ref{def-hc-App}) one obtains the 
correlation functions of the symmetric theory and in this case the 
critical point is uniquely defined by $h = 0$ by symmetry. The proper
definition of $r_c(u)$ requires more care, since we must perform a 
nonperturbative calculation in order to identify the critical point. 
For this purpose we will use the fact that in the critical crossover limit 
the perturbative expansion in powers of $u$ is equivalent to the 
perturbative expansion in the continuum $\phi^4$ theory once a proper 
mass renormalization is performed. 

In the continuum theory, if $\tilde{t}\equiv t_{\rm cont}/u_{\rm cont}$ 
is the adimensional reduced temperature defined so that $\tilde{t} = 0$ 
corresponds to the critical point, we have at one loop, 
cf. Eq.~(2.10) of Ref.~\cite{PRV-99}, 
\begin{equation}
u_{\rm cont} \chi_{2,\rm cont} = 
   {1\over \tilde{t}} + {1\over 8 \pi \tilde{t}^2} \left(
   \ln {8 \pi \tilde{t}\over 3} + 3 + 8 \pi D_2\right) + 
   O(\tilde{t}^{-3} \ln \tilde{t}^2),
\label{chi2-cont}
\end{equation}
where $D_2$ is a nonperturbative constant that can be expressed in terms 
of renormalization-group functions, cf.~Eq.~(2.11) of Ref.~\cite{PRV-99}. 
By using the four-loop perturbative results of Ref.~\cite{BNGM-77},
Ref.~\cite{PRV-99} obtained the estimate $D_2 = - 0.0524(2)$. 
It is not clear whether the error can really be trusted, since in two 
dimensions the resummation of the perturbative expansion is
not well behaved due to 
nonanalyticities of the renormalization-group functions at the 
fixed point \cite{CCCPV-00,CPV-01}; still, the estimate should provide the 
correct order of magnitude. 

The expansion \reff{chi2-cont} should be compared with the perturbative 
expansion of $\chi_2$ in the lattice model. We write $r_c(u)$ as 
\begin{equation}
r_c(u) = {u\over 8\pi} \ln u + {u\over2} \int_\mathbf{p}
      {V_3(\mathbf{p})^2\over K(\mathbf{p})^2} + A u ,
\label{defrc-const}
\end{equation}
where $A$ is a constant to be determined.
From Eq.~\reff{chi2-oneloop} we obtain
\begin{equation}
u \chi_2 = {u\over t} + {u^2\over 8 \pi t^2} 
   \left\{ \log{t\over 32 u} - 8 \pi A - 
          4\pi \int_\mathbf{p} \left[ 
          {V_4(\mathbf{p})\over K(\mathbf{p})} - 
          {1\over \hat{p}^2}\right]\right\} + 
          O(u^3).
\end{equation}
By comparing this result with Eq.~\reff{chi2-cont} we obtain
\begin{equation}
A = - D_2 - {1\over 8\pi} \ln {256\pi\over3} - {3\over 8\pi} - 
    {1\over2} \int_\mathbf{p} \left[
          {V_4(\mathbf{p})\over K(\mathbf{p})} - {1\over \hat{p}^2}\right].
\label{defA-const}
\end{equation}
If we use definition \reff{defrc-const} with $A$ fixed by 
Eq.~\reff{defA-const}, the critical point corresponds to $\tilde{t} = 0$.

\subsection{Critical crossover limit in the large-$N$ case} \label{sec6.6}

In Sec.~\ref{effHam-zeromode} we showed that the effective Hamiltonian
has the form \reff{Heff-ccl} with $u \sim 1/N$. However, in the 
large-$N$ case 
the parameters that can be tuned are $\Delta_m \equiv m_0^2 - m^2_{0c}$ and 
$\Delta_p \equiv p - p_c$, and moreover all quantities, beside $r$ and $H$,
depend on these two variables. We wish now to determine the changes, if any,
that appear in the previous treatment due to the dependence 
of $V^{(k)}(\mathbf{p}_1,\ldots, \mathbf{p}_k)$ and $K(\mathbf{p})$ 
on $\Delta_m$ and $\Delta_p$. 
As we show in the next Section, $\Delta_m$ and $\Delta_p$ scale 
respectively as $1/N^{1/3}$ 
and $1/N$, so that we can assume an additional dependence on $u^{1/3}$, 
i.e. we consider $K(\mathbf{p};u^{1/3})$ and 
$V^{(k)}(\mathbf{p}_1,\ldots, \mathbf{p}_k;u^{1/3})$. Note that, by definition,
$K(\mathbf{0};u^{1/3}) = 0$ for all values of $u$. 
Moreover, we assume, as in the case
of interest, that $V^{(3)}(\mathbf{0},\mathbf{0},\mathbf{0};u^{1/3}) = 0$ 
for all values of $u$. 
Following the calculation presented in Sec.~\ref{twoloop-ccl}, it is easy to 
realize that the dependence of the vertices on $u$ is irrelevant
except in $h_c(u)$. In this case, Eq.~\reff{chi1-oneloop} becomes
\begin{equation}
{t\over u} \chi_1 = - {h_c(u)\over u} - {1\over 2 \sqrt{u}} 
\int_\mathbf{p} {V_3(\mathbf{p};u^{1/3})\over K(\mathbf{p};u^{1/3}) + t} + 
O (\sqrt{u}).
\end{equation}
Because of the prefactor $1/\sqrt{u}$ we must take here into account the 
first correction proportional to $u^{1/3}$. Thus, if 
$V_3(\mathbf{p};u^{1/3}) \approx V_{3,0}(\mathbf{p}) + 
                        u^{1/3}  V_{3,1}(\mathbf{p})$ and 
$K(\mathbf{p};u^{1/3}) \approx K_{0}(\mathbf{p}) + u^{1/3}  K_{1}(\mathbf{p})$,
we obtain 
\begin{equation}
{t\over u} \chi_1 = - {h_c(u)\over u} - {1\over 2 \sqrt{u}} 
\int_\mathbf{p}
\left[{V_{3,0}(\mathbf{p}) + u^{1/3}  
       V_{3,1}(\mathbf{p})\over K_0(\mathbf{p}) + t} - 
       {u^{1/3} V_{3,0}(\mathbf{p}) K_1(\mathbf{p})\over 
         (K_0(\mathbf{p}) + t)^2}\right] + O(u^{1/6}).
\end{equation}
Now, $t$ can be set to zero without generating infrared divergences, 
neglecting corrections of order $u^{-1/2} t\ln t\sim u^{1/2} \ln u$.
It follows
\begin{equation}   
h_c(u) = - {\sqrt{u}\over 2} 
  \int_\mathbf{p} \left[ {V_{3,0}(\mathbf{p})\over K_0(\mathbf{p})} + 
    u^{1/3} {V_{3,1}(\mathbf{p}) K_0(\mathbf{p}) -  
             V_{3,0}(\mathbf{p}) K_1(\mathbf{p})\over K_0(\mathbf{p})^2}\right].
\label{hc-largeN}
\end{equation}
Eq.~\reff{hc-largeN} represents the only equation in which the explicit 
dependence of the vertices on $\Delta_m$ should be considered 
($\Delta_p$ is proportional to $u$ and thus it can always be set to zero).
In all other cases, we can simply set $\Delta_m = \Delta_p = 0$. 

\section{Crossover between mean-field and Ising behavior} \label{sec7}

In this Section we wish to apply the above-reported results
for the critical crossover limit to the Hamiltonian \reff{caleffchi} 
of the zero mode.
If we identify 
\begin{equation}
u \equiv {1\over N} 
  \bar{V}^{(4)}(\mathbf{0},\mathbf{0},\mathbf{0},\mathbf{0}),
\end{equation}
the effective Hamiltonian \reff{effexp2} 
corresponds to the Hamiltonian
discussed in Sec.~\ref{sec6}, apart from a rescaling of the 
three-leg and four-leg vertices.

We begin by defining the additive renormalization constants. 
The mass renormalization constant is given by, 
cf. Eqs.~\reff{defrc-const} and \reff{defA-const}, 
\begin{eqnarray}
r_c(N) &=& {u\over 8\pi} \ln \left( {3 u\over 256\pi}\right) - 
      {u\over 8\pi} (3 + 8 \pi D_2) 
\nonumber \\
   && + {u\over2} \int_\mathbf{p}\left\{
    {[\bar{V}^{(3)}(\mathbf{0},\mathbf{p},-\mathbf{p})]^2 \over 
     \bar{V}^{(4)}(\mathbf{0},\mathbf{0},\mathbf{0},\mathbf{0}) K(\mathbf{p})^2}
    -{\bar{V}^{(4)}(\mathbf{0},\mathbf{0},\mathbf{p},-\mathbf{p}) \over 
     \bar{V}^{(4)}(\mathbf{0},\mathbf{0},\mathbf{0},\mathbf{0}) K(\mathbf{p})} 
    + {1\over \hat{p}^2} \right\},
\label{rc1}
\end{eqnarray}
where $D_2$ is a nonperturbative constant defined in Ref.~\cite{PRV-99}
(numerically $D_2 \approx  - 0.052$). As discussed in Sec.~\ref{sec6.6}, 
we can compute all quantities appearing in Eq.~\reff{rc1} at the critical
point and keep only the leading terms for $N\to \infty$. We thus obtain
\begin{eqnarray}
r_c(N) &=& {u\over 8\pi} \ln \left( {3 u\over 256\pi}\right) -
      {u\over 8\pi} (3 + 8 \pi D_2)
\\
   && + {\alpha^2\over 2 N} 
    \int_\mathbf{p}\left\{
    [\tilde{V}^{(3)}_0(\mathbf{0},\mathbf{p},-\mathbf{p}) 
     \tilde{P}_0(\mathbf{p})]^2 - 
     \tilde{V}^{(4)}_0(\mathbf{0},\mathbf{0},\mathbf{p},-\mathbf{p}) 
     \tilde{P}_0(\mathbf{p}) + 
     {\alpha^2\over \hat{p}^2} 
     \tilde{V}^{(4)}_0(\mathbf{0},\mathbf{0},\mathbf{0},\mathbf{0}) \right\},
\nonumber 
\end{eqnarray}
where all quantities are computed at the critical point. 

Analogously, we should introduce a counterterm for the magnetic field: 
\begin{equation}
h_c(N) = - {1\over2} \sqrt{u} 
    \int_\mathbf{p} {\bar{V}^{(3)}(\mathbf{0},\mathbf{p},-\mathbf{p})\over 
    [\bar{V}^{(4)}(\mathbf{0},\mathbf{0},\mathbf{0},\mathbf{0})]^{1/2} 
     K(\mathbf{p})}.
\end{equation}
As discussed in Sec.~\ref{sec6.6}, such a quantity should not be simply 
computed at the critical point, but one should also take into account the 
additional corrections of order $N^{-1/3}$. Since $k\sim O(N^{1/6})$
in the critical crossover limit, we have 
\begin{eqnarray}
h_c(N) &=& - {\alpha\over 2\sqrt{N}} 
   \int_\mathbf{p} \left[ 
     \tilde{V}^{(3)}_0(\mathbf{0},\mathbf{p},-\mathbf{p}) - 
     {\tilde{V}^{(3)}_0(\mathbf{0},\mathbf{0},\mathbf{0})\over 
      \tilde{V}^{(4)}_0(\mathbf{0},\mathbf{0},\mathbf{0},\mathbf{0})} 
      \tilde{V}^{(4)}_0(\mathbf{0},\mathbf{0},\mathbf{p},-\mathbf{p})
     \right] {1\over \tilde{P}^{-1}_0(\mathbf{p}) - 
                \tilde{P}^{-1}_0(\mathbf{0}) } 
\nonumber \\
   && - {\alpha\over 2\sqrt{N}} 
     {\tilde{V}^{(3)}_0(\mathbf{0},\mathbf{0},\mathbf{0})\over
      \tilde{V}^{(4)}_0(\mathbf{0},\mathbf{0},\mathbf{0},\mathbf{0})}
      \int_\mathbf{p} 
    [\tilde{V}^{(3)}_0(\mathbf{0},\mathbf{p},-\mathbf{p}) 
     \tilde{P}_0(\mathbf{p})]^2 .
\end{eqnarray}
The second integral should be computed at the critical point, while the
first one and the prefactor of the second one should be expanded 
around the critical point. Since, as we shall show below, 
$\Delta_m \equiv m^2_0 - m_{0c}^2 \sim N^{-1/3}$ and $p - p_c \sim N^{-1}$, 
it is enough to compute the first correction in $\Delta_m$. 
In practice, we find that the renormalization terms have the form
\begin{eqnarray}
 r_c(N) &=& {1\over N} (r_{c0} \ln N + r_{c1}), \nonumber \\
 h_c(N) &=& {1\over \sqrt{N}} (h_{c0} + \Delta_m h_{c1}).
\label{rchc-exp}
\end{eqnarray}
Once $r_c(N)$ and $h_c(N)$ are computed, we can define the scaling variables
$x_t \sim t/u$ and $x_h \sim h/u$. Choosing the normalizations appropriately 
for later convenience, we define 
\begin{eqnarray}
x_t &\equiv& {N\over \alpha^2}  [\bar{P}^{-1}(0) - r_c(N)], 
\nonumber \\
x_h &\equiv& {N\over \alpha} [H - h_c(N)].
\label{xtxhdef}
\end{eqnarray}
Now, the critical crossover limit is obtained by tuning $p$ and $m_0^2$ 
around the critical point in such a way that $x_t$ and $x_h$ are 
kept constant, i.e. $p\to p_{\rm crit}$, $m_0^2\to m_{0,\rm crit}^2$, 
$N\to \infty$ at fixed $x_t$ and $x_h$. Note that here $p_{\rm crit}$ and 
$m_{0,\rm crit}^2$ correspond to the position of the critical point 
as a function of $N$ (thus $p_{\rm crit}\to p_c$ and 
$m_{0,\rm crit}^2\to m_{0c}^2$ as $N\to \infty$) and are obtaind by 
requiring $x_t=x_h=0$ (cf.~Sec.~\ref{sec6.5}). 

In order to compute the relation between $p$, $m_0^2$ and 
$x_t$, $x_h$ we set
$\Delta_m \equiv m^2_{0} - m^2_{0c}$ and 
$\Delta_p \equiv p - p_c$ and expand Eq.~\reff{xtxhdef} in 
powers of $\Delta_m$ and $\Delta_p$. In the following we shall show that 
$\Delta_m \sim N^{-1/3}$ and $\Delta_p\sim N^{-1}$,
so that the relevant terms are
\begin{eqnarray}
x_t &=& N (b_{t0} \Delta_p + b_{t1} \Delta_p \Delta_m + 
         b_{t2} \Delta_m^3) + d_1 \ln N + d_2 
\label{eqxt}
\\
x_h &=& N^{3/2} (b_{h0} \Delta_p^2 + b_{h1} \Delta_p \Delta_m + 
               b_{h2} \Delta_p \Delta_m^2 + b_{h3} \Delta_m^3 + 
               b_{h4} \Delta_m^4) + 
\nonumber \\ 
&& + N^{1/2} (d_{30} + d_{31} \Delta_m).
\label{eqxh}
\end{eqnarray}
The constants are obtained by expanding $\bar{P}^{-1}(\mathbf{0})$ and 
$H$ around the critical point and by using the expansions
\reff{rchc-exp}. All quantities are analytic in 
$\Delta_m$ and $\Delta_p$ and several terms are absent because of 
identities (\ref{id-P-CP}), (\ref{id-V3-CP}), and (\ref{id-V4-CP}). 
In particular, a term proportional to $\Delta_m^2$ is absent in the equation
for $x_t$. This is a consequence of Eq.~(\ref{id-V4-CP}). 
Indeed, we have
\begin{eqnarray}
{1\over \alpha^2} \bar{P}^{-1}(\mathbf{0}) &=& 
       \tilde{P}_0^{-1}(\mathbf{0}) - 
       {[\tilde{V}_0^{(3)}(\mathbf{0},\mathbf{0},\mathbf{0})]^2 \over 
        2 \tilde{V}_0^{(4)}(\mathbf{0},\mathbf{0},\mathbf{0},\mathbf{0})} 
      + O(\Delta_m^3,N^{-1}) 
\nonumber \\
   &=& {1\over 2} \left[
    {\partial^2 \tilde{P}_0^{-1}(\mathbf{0})\over \partial (m_0^2)^2} - 
     {1\over \tilde{V}_0^{(4)}(\mathbf{0},\mathbf{0},\mathbf{0},\mathbf{0})}
    \left( {\partial\over \partial m_0^2} 
     \tilde{V}_0^{(3)}(\mathbf{0},\mathbf{0},\mathbf{0}) \right)^2\right]
    \Delta_m^2 + O(\Delta_m^3,N^{-1}) 
\nonumber \\
    &=& 0 + 
    O(\Delta_m^3,N^{-1}),
\end{eqnarray}
where in the last step we have used Eq.~(\ref{id-V4-CP}).

We wish now to determine the behavior of $\Delta_m$ and $\Delta_p$ that is 
fixed by Eqs.~(\ref{eqxt}) and (\ref{eqxh}). We assume 
\begin{equation}
\Delta_m = {\delta_{m0}\over N^{\alpha}}, \qquad\qquad 
\Delta_p = {\delta_{p0}\over N^{\beta}}, 
\end{equation}
where $\delta_{m0}$ and $\delta_{p0}$ are nonvanishing constants and 
$\alpha$ and $\beta$ exponents to be determined. If 
$\beta < 1$, Eq.~\reff{eqxt} implies 
\be
b_{t0} \delta_{p0} N^{1-\beta} + b_{t2} \delta_{m0} N^{1-3\alpha} = 
  o(N^{1-\beta}),
\ee
which requires $\beta = 3\alpha$. Now, consider Eq.~(\ref{eqxh}).
The term $N^{3/2} \Delta_m^3$ is of order $N^{3/2 - \alpha}$ and cannot 
be made to vanish. Therefore, we must have $\beta \ge 1$. 
Considering again Eq.~(\ref{eqxh}), it is easy to see that all terms 
containing $\Delta_p$ cannot increase as fast as $N^{1/2}$. Therefore, 
cancellation of the term $d_{30} N^{1/2}$ requires $\alpha = 1/3$. 
Consideration of Eq.~\reff{eqxt} implies finally $\beta = 1$. 
This analysis can be extended to the subleading corrections, obtaining 
an expansion of the form
\begin{eqnarray}
\Delta_m &=& {\delta_{m0}\over N^{1/3}} + {\delta_{m1}\over N^{2/3}} + 
             {\delta_{m2}\over N^{5/6}},
           + O(N^{-1}) \nonumber \\
\Delta_p &=& {\delta_{p0}\over N} + 
             {\delta_{p1}\over N^{7/6}}
           + O(N^{-4/3}).
\end{eqnarray}
The coefficients are given by
\begin{eqnarray}
\delta_{m0} &=& - \left({d_{30}\over b_{h3}} \right)^{1/3} ,
\nonumber \\
\delta_{m1} &=& {\delta_{m0}^2 \over 3 b_{t0} d_{30}} 
   \left[ b_{t0} d_{31} - b_{h1} (d_1 \ln N + d_2) \right]
   + {b_{h1} \delta_{m0}^2\over 3 b_{t0} d_{30}} x_t + 
\nonumber \\
        && + {\delta_{m0}^5 \over 3 b_{t0} d_{30}}
   (b_{h4} b_{t0} - b_{h1} b_{t2}),
\nonumber \\ 
\delta_{p0} &=& - {1\over b_{t0}} \left( \delta_{m0}^3 b_{t2} +
             d_1 \ln N + d_2 - x_t\right).
\end{eqnarray}
We are not able to compute $\delta_{m2}$ and $\delta_{p1}$ but we can
however compute a relation between these two quantities. We obtain 
\begin{equation}
\delta_{p1} = - {3 \delta_{m0} \delta_{m2} b_{h3} \over b_{h1}} + 
       {1\over  b_{h1} \delta_{m0}} x_h.
\end{equation}
Correspondingly, by using Eq.~(\ref{expansion-gap}), we can compute the 
expansion of $u_h$:
\begin{eqnarray}
u_h &=& {u_{h0}\over N} + {u_{h1}\over N^{4/3}} + {u_{h2}\over N^{3/2}},
\end{eqnarray}
where $u_{h1}$ depends on $x_t$ and $u_{h2}$ on $x_h$. 
We thus define a new scaling field by requiring that no term proportional
to $x_t N^{-4/3}$ is present. For this purpose we set
\begin{equation}
\hat{u}_h = u_h + {x_{\rm mix} \over N^{1/3}} (p - p_c)  
\end{equation}
where the coefficient $x_{\rm mix}$ is given by
\begin{equation}
x_{\rm mix} = {(a_{03} b_{h1} - a_{11} b_{h3}) \delta_{m0} \over 
     b_{h3}}.
\end{equation}
The new scaling field has an expansion of the form 
\begin{equation}
\hat{u}_h= {\hat{u}_{h0}\over N} + {\hat{u}_{h1}\over N^{4/3}} + 
             {\hat{u}_{h2}\over N^{3/2}},
\end{equation}
where 
\begin{eqnarray}
\hat{u}_{h0} &= &
     a_{03} \delta_{m0}^3,
\\
\hat{u}_{h1} &=& {\delta_{m0}\over b_{h3}}
        [ - a_{03} d_{31} + 
          (a_{04} b_{h3} - a_{03} b_{h4}) \delta_{m0}^3  ]
\\
\hat{u}_{h2} &=& {a_{03}\over b_{h3} } x_h .
\end{eqnarray}
Interestingly enough, in $\hat{u}_{h2}$ all terms proportional to 
the unknown quantity $\delta_{m2}$ cancel.
At this point we can easily compute the $1/N$ expansion of the critical point
$p_{\rm crit}$, $\beta_{\rm crit}$.
It is enough to set $x_h$ = $x_t = 0$ in the previous expansions,
obtaining 
\begin{eqnarray}
   p_{\rm crit} &=& p_c + 
          \left. {\delta_{p0}\over N}\right|_{x_t = 0} + O(N^{-7/6}),
\\
  \beta_{\rm crit} &=& \beta_c 
         + {\hat{u}_{h0} +
            a_{10} \delta_{p0}|_{x_t = 0}\over N}  + O(N^{-7/6}).
\end{eqnarray}
It follows that 
\begin{eqnarray}
p - p_{\rm crit} &\approx &  \left(-  {2\over3} 
   {b_{t1} b_{h1} \delta_{m0}^2 \over d_{30}} + 1\right) {x_t\over b_{t0} N},
\label{eq:scaldp}
\\
\hat{u}_h - \hat{u}_{h,\rm crit} &\approx& 
           {3 a_{03} b_{t0} - a_{11} b_{t1} \over
                3 b_{h3} b_{t0} - b_{h1} b_{t1}}
           {x_h\over N^{3/2}},
\label{eq:scaluh}
\end{eqnarray}
where $\hat{u}_{h,\rm crit}$ is the value of $\hat{u}_h$ at the critical point.
Therefore, the $1/N$ corrections modify the position of the critical
point and change the magnetic scaling field which should now be identified with 
$\bar{u}_h = \hat{u}_h - \hat{u}_{h,\rm crit}$ (as we already discussed the 
thermal magnetic field is not uniquely defined and we take again
$p - p_{\rm crit}$). Eqs.~(\ref{eq:scaldp}) and (\ref{eq:scaluh}) also indicate
which are the correct scaling variables. Ising behavior is observed only if 
$N(p - p_{\rm crit})$ and $N^{3/2}(\hat{u}_h - \hat{u}_{h,\rm crit} )$ 
are both small; in the opposite case mean-field behavior is observed.
Therefore, as $N$ increases the width of the critical region 
decreases, and no Ising behavior is observed at $N=\infty$ exactly. 

\section{Critical behavior of $\langle\bsigma_x\cdot \bsigma_{x+\mu}\rangle$
} \label{sec8}

In this Section we wish to compute the large-$N$ behavior of 
\begin{equation}
 \overline{E} = \< \bsigma_x \cdot \bsigma_{x+\mu}\> .
\end{equation}
Such a quantity does not coincide with the energy. 
However, as far as the critical behavior is concerned, there should not be 
any significant difference. In order to perform the computation, 
note that the equations of motion for the field $\lambda_{x\mu}$ give
\begin{equation}
\< \rho_{x\mu} \> = 1 + \overline{E }.
\end{equation}
Thus, we have
\begin{equation}     
\overline{E} = \tau - 1 + {1\over \sqrt{N}} \< \hat{\rho}_{x\mu} \> = 
    \tau - 1 + {1\over \sqrt{N}} \sum_B U_{4B}(\mathbf{0}) 
    \< \Phi_{Bx} \>.
\end{equation}
Therefore, we need to compute the correlations $\<\phi\>$ and 
$\<\varphi_a\>$. For the zero mode we have immediately 
\be
\< \phi_x\> = \alpha \< \chi_x \> - k = 
   {\alpha \over x_t} f_1^{\rm symm}(x_t,x_h) - k,
\ee
where $f_1^{\rm symm}(x_t,x_h)$ is the crossover function for the 
magnetization in the Ising model and the constant $k$ diverges as $N^{1/6}$, modulo some obvious normalizations.

Let us now consider $\<\varphi_a \>$ and show that such a correlation vanishes
as $N\to\infty$ in the critical crossover limit. Indeed, we can write
\begin{eqnarray}
\<\varphi_a \> &=& {1\over \sqrt{N}} 
 \hat{P}_{ab}(\mathbf{0}) 
   \int_{\mathbf{p}} [\hat{V}_{bcd}(\mathbf{0},\mathbf{p},-\mathbf{p})
             \hat{P}_{cd}(\mathbf{p}) + 
      (\hat{V}_{b11}(\mathbf{0},\mathbf{p},-\mathbf{p}) - 
      \hat{V}_{b11}(\mathbf{0},\mathbf{0},\mathbf{0})) \hat{P}_{11}(\mathbf{p})]
\nonumber \\
       && + {1\over \sqrt{N}} \hat{P}_{ab}(\mathbf{0}) 
            \hat{V}_{b11}(\mathbf{0},\mathbf{0},\mathbf{0}) \<\phi^2_x\>.
\end{eqnarray}
Note that in the last term we have replaced 
\begin{equation}
\int_\mathbf{p} \hat{P}_{11}(\mathbf{p}) = \<\phi^2_x\>,
\end{equation}
a necessary step, since the integral diverges at the critical point and 
therefore should be computed in the effective theory for the zero mode. 
Now, we have 
\begin{equation}
\<\phi^2_x\>  = k^2 - 2 \alpha k \< \chi_x \> + \alpha^2 \< \chi_x^2 \>.
\end{equation}
In the critical crossover limit, the two expectation values are replaced 
by the crossover functions for the magnetization and the energy and by 
a regular term. Therefore, for $N\to \infty$, the leading behavior is 
$\<\phi^2_x\>  = k^2 \sim N^{1/3}$. It follows that 
$\<\varphi_a\> \sim N^{-1/6}$. In conclusion we can write
\begin{eqnarray}
\overline{E} &= &
    \tau - 1 + {1\over \sqrt{N}} U_{41}(\mathbf{0}) 
    \left[
   {\alpha\over x_t} f_1^{\rm symm}(x_t,x_h) - k \right]
\nonumber \\
 &=& \overline{E}_{\rm reg} + {1\over \sqrt{N}} U_{41}(\mathbf{0})
     {\alpha\over x_t} f_1^{\rm symm}(x_t,x_h) + O(N^{-2/3}),
\label{E-largeN}
\end{eqnarray}
where $\overline{E}_{\rm reg}$ is the regular part of $\overline{E}$:
\begin{eqnarray} 
  \overline{E}_{\rm reg} &= & \tau - 1 - {1\over \sqrt{N}} U_{41}(\mathbf{0}) k
\nonumber \\
   &=& \tau_0 - 1 + [\tau_1 - k_{0} U_{41}(\mathbf{0})] \delta_{m0} 
      N^{-1/3} + O(N^{-2/3}),
\end{eqnarray}
where we have written $\tau \approx \tau_0 + \tau_1 \Delta_m$ and 
$k \approx k_{0} \Delta_m \sqrt{N}$. 

Equation \reff{E-largeN} shows that the singular part of $\overline{E}$ 
behaves as the magnetization
in the Ising model. For $x_t \ll 1$ and $x_h\ll 1$ one observes Ising behavior
and thus 
\begin{eqnarray}
\overline{E} - \overline{E}_{\rm reg} \sim |x_t|^{\beta_I} && \qquad \hbox{for}\quad x_h = 0, 
      \hbox{low $t$ phase}
\nonumber \\
\overline{E} - \overline{E}_{\rm reg} \sim |x_h|^{1/\delta_I} && \qquad \hbox{for}\quad x_h \not= 0,
\end{eqnarray}
where $\beta_I = 1/8$ and $\delta_I = 15$. On the other hand, in the opposite
limit we have
\begin{eqnarray}
\overline{E} - \overline{E}_{\rm reg} \sim |x_t|^{\beta_{MF}} && \qquad \hbox{for}\quad x_h = 0, 
           |x_t|\to \infty,
      \hbox{low $t$ phase}
\nonumber \\
\overline{E} - \overline{E}_{\rm reg} \sim |x_h|^{1/\delta_{MF}} && \qquad \hbox{for}\quad 
           |x_h|\to \infty
\end{eqnarray}
with $\beta_{MF} = 1/2$ and $\delta_{MF} = 3$. Note that the limit 
$|x_t|\to \infty$ and $|x_h|\to\infty$ should always be taken close to the 
critical limit. Therefore, $|x_h|\to\infty$ means that we should consider 
$N\to \infty$, $\hat{u}_h\to \hat{u}_{h,\rm crit}$, $p\to p_{\rm crit}$
in such a way that $N^{3/2}(\hat{u}_h - \hat{u}_{h,\rm crit})\to \infty$,
i.e. $N$ should increase much faster that the rate of approach to the 
critical point.

\section{Numerical results for selected Hamiltonians} \label{sec9}

In this Section we present some numerical results for some selected 
Hamiltonians. First,
as in Ref. \cite{BGH-02}, we consider 
\be
W(x) = {2\over p}\left( {x\over2}\right)^p.
\label{Ham1}
\ee
Second, we consider the mixed O($N$)-RP$^{N-1}$ model with 
Hamiltonian \cite{MR-87} 
\be
{\cal H} = - N \beta_V \sum_{x\mu} (\bsigma_x\cdot \bsigma_{x+\mu}) -
{N\beta_T\over2} \sum_{x\mu} (\bsigma_x\cdot \bsigma_{x+\mu})^2.
\ee
This Hamiltonian corresponds to the function 
\be
W(x) = p x + {1\over 4} (1 - p) x^2,
\label{Ham2}
\ee
where we set $\beta_V = (1 + p) \beta/2$ and $\beta_T = (1 - p) \beta/2$. 
This Hamiltonian is ferromagnetic for $p > -1$. 
Note that for $p = -1$ we obtain the $RP^{N-1}$ Hamiltonian 
\cite{HM-82,MR-87,Ohno_90,KZ,Butera_92,CEPS_RP,Hasenbusch_96,%
Niedermayer_96,Catterall_98,SS-01}
\be
{\cal H} = - {N \beta \over 2} \sum_{x\mu} (\bsigma_x\cdot\bsigma_{x+\mu})^2,
\ee
that has the additional gauge invariance $\bsigma_x \to \epsilon_x \bsigma_x$,
$\epsilon_x = \pm 1$. Under standard assumptions, the large-$N$
analysis should apply also to this last model: the local gauge invariance 
should not play any role.\footnote{
The irrelevance of the $\mathbb{Z}_2$ symmetry for the large-$\beta$ 
behavior of $RP^{N-1}$ models have been discussed in detail in 
Ref.~\cite{Hasenbusch_96,Catterall_98,Niedermayer_96}. Thus, in spite of 
the additional local invariance, $RP^{N-1}$ models are 
expected to be asymptotically free and to be described by the 
perturbative renormalization group.}

In the large-$N$ limit \cite{CP-02}, the first Hamiltonian has a critical 
point at $\beta_c\approx 1.335 $ and $p_c \approx 4.538 $. 
By using the numerical results reported in Table~\ref{numerical}
we can compute the first corrections to the critical parameters. 
We obtain
\begin{eqnarray}
\beta_c  &\approx&  1.335 + {1\over N} \left( 36.127 +2.093 \ln N \right),
\\
p_c  &\approx& 4.538 + {1\over N}\left(87.92+4.80 \ln N \right).
\label{pc-W1}
\end{eqnarray}
Note the presence of a $\ln N/N$ correction due to the nonanalytic 
nature of the renormalization counterterms. The correction terms are 
quite large, indicating that the large-$N$ results are
quantitatively predictive only for large values of $N$. This is not 
totally unexpected since \cite{BGH-02} $p_c \approx 20$ for $N=3$, 
that is quite far from the large-$N$ estimate $p_c \approx 4.538$. 
If we substitute $N=3$ in Eq.~\reff{pc-W1}, we obtain 
$p_c \approx 35$, which shows that the corrections have the 
correct sign and give at least the correct order of magnitude. 

\begin{table}[t]
\begin{center}
\begin{tabular}{ccc}
\hline\hline
& ${\cal H}_1$ & ${\cal H}_2$ \\
\hline
$\beta_c$   &     1.334721915850    &     0.9181906464057     \\
$p_c$       &     4.537856778637    &  $-$0.9707166650184     \\
$m_{0c}^2$  &     0.1501849439193   &     0.8657494320430     \\
$a_{10} $   &     0.4359516292302   &  $-$1.1359750388653     \\
$a_{11} $   &     0.5042522341176   &  $-$0.6248171602172     \\
$a_{12} $   &  $-$1.2793594495686   &     0.1122363376128     \\
$a_{03} $   &  $-$1.3015714087645   &  $-$0.0061080440602     \\
$a_{04} $   &     67.512019516378   &     0.2847340288532     \\
$b_{t0} $   &  $-$0.04261881236908  &     0.21169346791995    \\
$b_{t1} $   &  $-$0.0043954143923   &     0.05334711842197    \\
$b_{t2} $   &     1.85335235385232  &     0.00015229016444    \\
$b_{h0} $   &  $-$0.0077917231312   &  $-$0.664558775414698   \\
$b_{h1} $   &     0.085888253027    &  $-$0.094120307536470   \\
$b_{h2} $   &  $-$0.5785580673083   &     0.049596987743194   \\
$b_{h3} $   &  $-$0.221693999401    &  $-$0.000920094744508   \\
$d_1    $   &     0.205             &     0.2678             \\
$d_2    $   &     0.614             &  $-$0.808514            \\
$d_{30} $   &     0.374             &     1.1692              \\
$\alpha^{-2}$ &     0.0315969         &     0.00933367          \\
\hline\hline
\end{tabular}
\end{center}
\caption{Numerical estimates for Hamiltonian \reff{Ham1} (${\cal H}_1$)
and \reff{Ham2} (${\cal H}_2$).}
\label{numerical}
\end{table}

For $N=\infty$ the second Hamiltonian has a critical point
at \cite{MR-87} $\beta_c\approx 0.918 $ and $p_c \approx  -0.971 $. 
Including the first correction we obtain
\begin{eqnarray}
\beta_c  &=& 0.9182-{1\over N}\left(11.062-1.437 \ln N\right),
\\
p_c & =& -0.9707 +{1\over N}\left( 2.905 - 1.265\ln N \right).
\end{eqnarray}
In this case the corrections are smaller. However, for 
$N \le 99.4$ they predict $p_c < -1$, although only slightly. 
This is of course not possible, since for $p < -1$ the system 
is no longer ferromagnetic. Thus, we expect the transition to disappear 
for some $N = N_c$, with $N_c \approx 100$ (this is of course a very 
rough estimate). Since $p_c \approx - 1$, we also predict 
$RP^{N-1}$ models to show a very weak first-order transition for large
$N$.

\section{Conclusions} \label{sec10}

In this paper we investigate generic one-parameter families 
of $N$-vector models that show a line of first-order 
finite-$\beta$ transitions. We focus on the endpoint of the 
first-order transition line where energy-energy correlations 
become critical, while the spin-spin correlation length 
remains finite, in agreement with Mermin-Wagner theorem \cite{MW-66}. 
We show that, at the critical point, the standard $1/N$ expansion
breaks down, since the inverse propagator of the auxiliary fields 
has a zero eigenvalue. A careful treatment shows that the zero mode,
i.e. the field associated with the vanishing eigenvalue, has an effective 
Hamiltonian that corresponds to a weakly coupled one-component $\phi^4$ theory.
Thus, the phase transition belongs to the Ising universality class for any
$N$, in agreement with the argument of Ref.~\cite{BGH-02}. In Ref.~\cite{CP-02}
it was shown that for $N=\infty$ the transition has mean-field exponents.
We reconcile here these two results. If $u_t$ and $u_h$ are 
the linear thermal and magnetic scaling fields, in the critical limit 
a generic long-distance quantity $\cal O$
has a behavior of the form
\be
\langle {\cal O} \rangle_{\rm sing} \approx 
  u_t^\sigma f_{\cal O}(u_t N, u_h N^{3/2}),
\ee
where $f(x,y)$ is a crossover function. Only if 
$u_t N \ll 1$ and $u_h N^{3/2} \ll 1$ does one observe Ising behavior.
In the opposite limit one observes mean-field criticality. 
Therefore, the width of the Ising critical region goes to zero as $N\to\infty$
and, even if the transition is an Ising one for any $N$, only 
mean-field behavior is observed for $N=\infty$. The behavior observed 
at the critical point for $N\to\infty$ rensembles very closely what 
is observed in medium-range models 
\cite{crossover1,crossover2,crossover3,crossover4}, with $N$ 
playing the role of the interaction range. Our analysis fully confirms
the conclusions of Ref.~\cite{BGH-02}.

From a more quantitative point of view, we give explicit expressions 
for the critical values $\beta_c$ and $p_c$ and for the nonlinear scaling 
fields to order $1/N$. Numerical results are given for the 
Hamiltonian introduced in Ref.~\cite{BGH-02} and for mixed 
$O(N)$-$RP^{N-1}$ models \cite{MR-87}.

Finally, note that, even though the model we consider is two-dimensional,
the same discussion can be done in generic dimension $d$. Indeed,
for any $d$ the zero mode has a Hamiltonian that corresponds to 
a weakly coupled $\phi^4$ theory. Thus, for $d < 4$, the phase transition
is always Ising-like. Some changes, however, should be done in the scaling
equations. For generic $d < 4$, Eq.~\reff{scaling-cross-lim} should be 
replaced by 
\be
\chi_n = u^{d/(4-d)} t^{-n(d+2)/4}
   f_n^{\rm symm}(u t^{-(4-d)/2}, u h^{-2(4-d)/(2+d)}).
\ee
Also the expressions for the renormalization constants 
$r_c$ and $h_c$ should be changed: for instance, in three dimensions 
we also expect contributions from two-loop graphs, as it happens in
medium-range models \cite{PRV-99}. The main result of this paper, i.e. the fact
that the width of the Ising critical region goes to zero as $N\to\infty$, 
holds in any $d < 4$. However, in generic dimension $d$ the natural 
scaling variables are $(p - p_{\rm crit}) N^{2/(4-d)}$ and 
$(\hat{u}_h - \hat{u}_{h,\rm crit}) N^{3/(4-d)}$. Thus, for $d> 2$ the 
Ising critical region shrinks faster as $N$ increases: in three 
dimensions as $N^{-2}$ and $N^{-3}$ in the thermal and magnetic directions
respectively.

\appendix

\section{Identities among effective vertices} \label{AppA}

In this Section we wish to derive a general set of identities 
for the zero-momentum vertices. For this purpose we first rewrite
the propagator and the vertices by considering the constants 
$\alpha$, $\gamma$, and $\tau$ appearing in Eq.~\reff{field-exp} 
as independent variables {\em without} assuming the saddle-point relations. 
We define the integrals 
\begin{equation}
{\cal A}_{ij,n}(\alpha,\gamma) = 
  \int_\mathbf{q} 
    {\cos^i q_x \cos^j q_y\over (\gamma - \alpha \sum_\mu \cos q_\mu)^n}.
\end{equation}
It is easy to verify that, if $\alpha$ and $\gamma$ are replaced by 
their saddle-point values, $\alpha = 2 W'(\bar{\tau})$ and 
$\gamma = (4 + m_0^2) W'(\bar{\tau})$, then 
\begin{equation}
{\cal A}_{ij,n}(\alpha,\gamma) = 
  {1\over [W'(\overline{\tau})]^n} 
   \int_\mathbf{q}
 {\cos^i q_x \cos^j q_y\over (\hat{q}^2 + m_0^2)^n}.
\end{equation}
In terms of $\alpha$, $\gamma$, and $\tau$ the propagator is simply obtained
by using Eq.~\reff{propagator}
and replacing $A_{i,j}(0,m_0^2)/[W'(\overline{\tau})]^2$ with 
${\cal A}_{ij,2}(\alpha,\gamma)$. Let us now consider a generic 
$n$-leg vertex $V^{(n)}_{A_1,\ldots,A_n}$ at zero momentum. It is easy to 
verify that the only nonvanishing terms with $A_i = 4$ or $A_i = 5$ 
for some $i$ are (in this Section we do not explicitly write the momentum 
dependence since in all cases we are referring to zero-momentum 
quantities)
\begin{equation}
V^{(n)}_{4,\ldots 4} = V^{(n)}_{5,\ldots 5} = - \beta W^{(n)}(\tau)\; .
\end{equation}
If all $A_i \le 3$, Eq.~\reff{defVn} gives
\begin{equation}
V^{(n)}_{A_1,\ldots,A_n} = {1\over2} (-1)^{i_2+i_3+n+1} (n-1)! 
    {\cal A}_{i_2 i_3,n}(\alpha,\gamma),
\end{equation}
where $i_2$ (resp. $i_3$) is the number of indices equal to 2 (resp. to 3).
These expressions allow us to obtain simple recursion relations for the 
vertices. 
\begin{eqnarray}
&& V^{(n+1)}_{1,A_1,\ldots,A_n} = 
   {\partial\over \partial\gamma} V^{(n)}_{A_1,\ldots,A_n},
\nonumber \\
&& V^{(n+1)}_{2,A_1,\ldots,A_n}  + V^{(n+1)}_{3,A_1,\ldots,A_n} = 
   {\partial\over \partial\alpha} V^{(n)}_{A_1,\ldots,A_n}, 
\nonumber \\  
&& V^{(n+1)}_{4,4,\ldots,4}  = 
   {\partial\over \partial\tau} V^{(n)}_{4,\ldots,4},
\nonumber \\  
&& V^{(n+1)}_{5,5,\ldots,5}  = 
   {\partial\over \partial\tau} V^{(n)}_{5,\ldots,5},
\end{eqnarray}
where $\alpha$, $\gamma$, $\tau$, and $\beta$ are considered as independent 
variables.
These relations also apply to the case $n = 2$, once we identify 
$V^{(2)}_{AB} = P^{-1}_{AB}$. 

Let us consider the projector on the zero mode $v_A$, cf.~Eq.~\reff{depPhi}. 
Keeping into account that the symmetry of the matrix at zero momentum 
implies $v_2 = v_3$ and $v_4 = v_5$, we obtain 
\begin{eqnarray}
\sum_B v_B V^{(n+1)}_{B,A_1,\ldots,A_n} = 
   \left( v_1 {\partial\over \partial\gamma} + 
          v_2 {\partial\over \partial\alpha} + 
          v_4 {\partial\over \partial\tau} \right)
    V^{(n)}_{A_1,\ldots,A_n}.
\end{eqnarray}
Close to the critical point we can write, cf. Eq.~\reff{defz},
\begin{equation}
  v_A = \hat{z}_A + O(s_1)
\end{equation}
with
\begin{eqnarray}
   \hat{z}_A = K \left( 2 {{\cal A}_{01,2}\over {\cal A}_{00,2}}, 1, 1,
   {1\over 2 W''(\tau)}, {1\over 2 W''(\tau)} \right),
\end{eqnarray}
where $K$ is such to have $\sum_B \hat{z}_B^2 = 1$. 
Now, let us note that at the saddle point we have
(we now think of $\alpha$, $\tau$, and 
$\gamma$ as functions of $m_0^2$ and $p$)
\begin{eqnarray}
&& {\partial \gamma\over \partial m_0^2} = 
   W'' {\partial\tau\over \partial m_0^2} 
     \left(4 + m_0^2 - {B_1\over B_2}\right) + O(s_1) = 
   {2 W''\over K} {\partial \tau\over \partial m_0^2} \hat{z}_1 + O(s_1),
\nonumber \\
&& {\partial \alpha\over \partial m_0^2} = 
   2 W'' {\partial \tau\over \partial m_0^2} = 
   {2 W''\over K}  {\partial \tau\over \partial m_0^2} \hat{z}_2 + O(s_1),
\end{eqnarray}
where we have used Eqs.~\reff{dbdm} and \reff{relAijBn}.
Thus, if we define 
\begin{equation}
   C \equiv \left( {2 W''\over K}  {\partial \tau\over \partial m_0^2}
            \right)^{-1},
\end{equation}
we can rewrite
\begin{eqnarray}
\sum_B v_B V^{(n+1)}_{B,A_1,\ldots,A_n} &=& C
   \left( {\partial \gamma\over \partial m_0^2}
          {\partial\over \partial\gamma} + 
          {\partial \alpha\over \partial m_0^2}{\partial\over \partial\alpha} + 
         {\partial \tau\over \partial m_0^2}{\partial\over \partial\tau} 
         \right) 
    V^{(n)}_{A_1,\ldots,A_n}
    + O(s_1) 
\nonumber \\ 
   &= &
    C {\partial\over \partial m_0^2} 
    V^{(n)}_{A_1,\ldots,A_n} + O(s_1),
\label{identity1}
\end{eqnarray}
where in the last term $\alpha$, $\gamma$, $\tau$, and $\beta$ take their
saddle-point values in terms of $m_0^2$ and $p$.

To go further we compute the derivative of $\hat{z}_A$ with respect to $m_0^2$. 
Since 
\begin{equation}
   \sum_B P^{-1}_{AB} \hat{z}_B = O(s_1),
\end{equation}
we have 
\begin{equation}
\sum_B P^{-1}_{AB} {\partial \hat{z}_B\over \partial m_0^2} = 
    - \sum_B {\partial P^{-1}_{AB} \over \partial m_0^2} \hat{z}_B + O(s_2),
\end{equation}
where $s_2 = \partial^2 \beta / \partial (m_0^2)^2$ also vanishes at 
the critical point. 

Since $\sum_A \hat{z}_A^2 = 1$, $\partial \hat{z}_B/ \partial m_0^2$ 
belongs to the subspace orthogonal to $\hat{z}_A$. 
Thus, if $P^\bot_{AB}$ is the inverse of $P^{-1}$ in the massive subspace,
we obtain
\begin{equation}
 {\partial \hat{z}_A\over \partial m_0^2} 
  = - \sum_{BC} P^\bot_{AB} {\partial P^{-1}_{BC} \over \partial m_0^2} 
   \hat{z}_C + O(s_2) = 
  - {1\over C} \sum_{BCD} P^\bot_{AB}  V^{(3)}_{BCD} \hat{z}_C \hat{z}_D + 
    O(s_2).
\label{dhatz}
\end{equation}
Finally, let us compute 
\begin{eqnarray}
\hat{V}_{1,\ldots 1}^{(n+1)} &=& C 
   \sum_{A_1,\ldots,A_n} z_{A_1} \ldots z_{A_n} 
     {\partial \over \partial m_0^2} 
       V^{(n)}_{A_1,\ldots,A_n}  + O(s_1)
\nonumber  \\
   & = & C {\partial \over \partial m_0^2}  \hat{V}_{1,\ldots 1}^{(n)} - 
      n C \sum_{A_1,\ldots,A_n} z_{A_1} \ldots z_{A_{n-1}} 
          V^{(n)}_{A_1,\ldots,A_n} {\partial z_{A_n}\over \partial m_0^2} 
       + O(s_1)
\nonumber  \\
   & = & C {\partial \over \partial m_0^2}  \hat{V}_{1,\ldots 1}^{(n)}
     + n \sum_{ab} \hat{V}_{11a}^{(3)} \hat{P}_{ab} 
          \hat{V}_{b1\ldots 1}^{(n)} + O(s_2),
\label{identity2}
\end{eqnarray}
where the indices $a$ and $b$ run from 1 to 4 over the massive subspace.
Since $s_2$ is of order $p-p_c$ and $m_0^2 - m_{0c}^2$ this relation 
implies an identity only at the critical point. 
However, for $n=2$ we can obtain another relation.
Since $V^{(2)}_{AB} = P^{-1}_{AB}$ and $\sum_B P^{-1}_{AB} z_B = O(s_1)$
we do not need Eq.~\reff{dhatz} and obtain directly
\begin{eqnarray}
\hat{V}_{111} = C {\partial \over \partial m_0^2} P^{-1}_{11} + O(s_1).
\label{identity3}
\end{eqnarray}
The presence of corrections of order $s_1$ gives rise 
to two critical-point identities. Since $P^{-1}_{11}\sim s_1$ we obtain
\begin{eqnarray}
 \hat{V}_{111} & = & 0, \nonumber \\
 {\partial \over \partial m_0^2} \hat{V}_{111} & = & 
    C {\partial^2 \over \partial (m_0^2)^2} P^{-1}_{11} ,
\end{eqnarray}
where all quantities are computed at the critical point.

\end{document}